\documentclass[conference]{IEEEtran}

\usepackage[utf8]{inputenc}
\usepackage[T1]{fontenc}
\usepackage{microtype}

\usepackage{graphicx}
\DeclareGraphicsExtensions{.pdf}
\usepackage{xcolor}
\usepackage{algorithm}
\usepackage{algpseudocodex}
\usepackage{amsmath}
\usepackage{amssymb,amsfonts}
\usepackage{listings}
\usepackage{xspace}
\usepackage{multirow}
\usepackage{semantic} %
\usepackage{tikz}
\usetikzlibrary{arrows.meta} %
\usepackage{mathtools} %
\usepackage{siunitx}
\usepackage{placeins}

\usepackage[space]{cite}
\renewcommand\citedash{{\kern 0.16em}--{\kern 0.05em}}

\usepackage{textcomp}
\usepackage{booktabs}
\usepackage{tablefootnote}
\usepackage{enumitem} %

\usepackage[hidelinks]{hyperref}

\algrenewcommand\algorithmicrequire{\textbf{Input:}}
\DeclareMathOperator{\Sol}{Sol}
\DeclareMathOperator{\Func}{Func}
\DeclareMathOperator{\CallFunc}{Call}
\DeclareMathOperator{\ImportedFunc}{ImpFunc}
\newcommand{\PowerSet}{\mathcal{P}}
\newcommand{\external}{\ensuremath\Omega\xspace}
\newcommand{\implicit}{IP\xspace}
\newcommand{\explicit}{EP\xspace}
\newcommand{\numNonEmptyC}{3\,659\xspace}
\newcommand{\numConfigs}{208\xspace}

\newcommand{\customarrow}[1]{
\tikz[baseline=-\the\dimexpr\fontdimen22\textfont2\relax]{
\node[anchor=south,font=\scriptsize, inner ysep=1.5pt,outer xsep=2.2pt, text width=5mm](x){};
\draw[shorten <=3.6pt,shorten >=3.6pt,>={latex[scale=1.4]},thick,#1](x.south west)--(x.south east);
}
}
\newcommand{\simplearrow}{\customarrow{->}}
\newcommand{\loadarrow}{\customarrow{arrows={Circle[fill=white]->},dash pattern=on 2pt off 2pt}}
\newcommand{\storearrow}{\customarrow{arrows={-Circle[fill=white]>},dash pattern=on 2pt off 2pt}}
\newcommand*\circled[1]{\tikz[baseline=(char.base)]{
            \node[shape=circle,draw,inner sep=0,minimum size=11pt] (char) {#1};}}

\newcommand{\ie}{i.e.\xspace}
\newcommand{\eg}{e.g.\xspace}
\newcommand{\etal}{et al.\xspace}
\newcommand{\etc}{etc.\xspace}

\newif\ifaam
\aamtrue

\newif\ifnva
\nvafalse

\ifaam
    \usepackage[absolute,overlay]{textpos}
    
    \newcommand\copyrighttext{%
      \footnotesize \textcopyright 2025 IEEE. Personal use of this material is permitted.
      Permission from IEEE must be obtained for all other uses, in any current or future
      media, including reprinting/republishing this material for advertising or promotional
      purposes, creating new collective works, for resale or redistribution to servers or
      lists, or reuse of any copyrighted component of this work in other works.}
    \newcommand\copyrightnotice{%
    \begin{tikzpicture}[remember picture,overlay]
    \node[anchor=south,yshift=20pt] at (current page.south)
      {\fbox{\parbox{\dimexpr\textwidth-\fboxsep-\fboxrule\relax}{\copyrighttext}}};
    \end{tikzpicture}%
    }
\fi

\begin{document}

\title{PIP: Making Andersen's Points-to Analysis Sound and Practical for Incomplete C Programs}

\author{%
\IEEEauthorblockN{
Håvard Rognebakke Krogstie}
\IEEEauthorblockA{
\textit{Norwegian University of Science and Technology} \\
Trondheim, Norway\\
havard.r.krogstie@ntnu.no
}
\\
\IEEEauthorblockN{%
Magnus Själander}
\IEEEauthorblockA{
\textit{Norwegian University of Science and Technology} \\
Trondheim, Norway \\
magnus.sjalander@ntnu.no
}
\and
\IEEEauthorblockN{%
Helge Bahmann}
\IEEEauthorblockA{
\textit{Independent Researcher} \\
Zürich, Switzerland \\
hcb@chaoticmind.net
}
\\
\IEEEauthorblockN{%
Nico Reissmann}
\IEEEauthorblockA{
\textit{Independent Researcher} \\
Trondheim, Norway \\
nico.reissmann@gmail.com
}
}

\maketitle
\ifaam
    \copyrightnotice
    \thispagestyle{empty}
    \pagestyle{plain}
\fi

\begin{abstract}
Compiling files individually lends itself well to parallelization, but forces the compiler to operate on incomplete programs.
State-of-the-art points-to analyses guarantee sound solutions only for complete programs,
requiring summary functions to describe any missing program parts.
Summary functions are rarely available in production compilers, however, where soundness and efficiency are non-negotiable.
This paper presents an Andersen-style points-to analysis that efficiently produces sound solutions for incomplete C programs.
The analysis accomplishes soundness by tracking memory locations and pointers that are accessible from external modules, and efficiency by performing this tracking implicitly in the constraint graph.
We show that implicit pointee tracking makes the constraint solver 15$\times$ 
faster than any combination of five different state-of-the-art techniques using explicit pointee tracking.
We also present the Prefer Implicit Pointees (PIP) technique that further reduces the use of explicit pointees.
PIP gives an additional speedup of 1.9$\times$, compared to the fastest solver configuration not benefiting from PIP.
The precision of the analysis is evaluated in terms of an alias-analysis client, where it reduces the number of \texttt{MayAlias}-responses by 40\% compared to LLVM's BasicAA pass alone.
Finally, we show that the analysis is scalable in terms of memory,
making it suitable for optimizing compilers in practice.
\end{abstract}
\vspace{1mm}
\begin{IEEEkeywords}
Static analysis, points-to analysis, partial analysis.
\end{IEEEkeywords}

\section{Introduction}
\label{sec:intro}
Alias information is a prerequisite for program analyses and transformations that are essential in optimizing compilers, program verification, and program comprehension tools.
Important compiler transformations, such as loop invariant code motion, dead load and store elimination~\cite{Surendran:CC2014}, loop versioning~\cite{Chitre:OOPSLA2022}, vectorization~\cite{Karrenberg:CGO2011}, \etc, rely on alias information to be effective.
A points-to analysis provides sets of possible targets for pointers, and can form the basis for an alias analysis, in addition to other analyses like call graph and mod/ref summary creation~\cite{Lattner:PLDI2007}.
Ideally, a points-to analysis should \emph{efficiently} provide \emph{precise} information,
but a balance must be found in practice, as precise points-to analysis is undecidable~\cite{Landi:LOPLAS92}.
Practical and scalable implementations therefore approximate the exact solution, and a plethora of trade-offs between precision and performance exist~\cite{Lei:SAS2019, Ye:SAS2014, Sui:CGO2013, Nasre:CGO2011, Yu:CGO2010, Pereira:CGO2009, Lattner:PLDI2007, Zhu:DAC2005}.

These trade-offs receive a lot of attention in the literature, where their scalability is shown by performing whole-program analysis.
However, even if all optimizations are deferred to link time, which is the exception rather than the norm, very few programs outside the domain of embedded systems are complete.
On the contrary, the majority of programs are incomplete, \ie, programs with unknown external functions,
as they depend on external libraries and/or system calls~\cite{Hind:PASTE2001, Lattner:PLDI2007}.

\begin{figure}%
{\footnotesize
\begin{lstlisting}[language=C,,numbers=left,numbersep=-6pt]
   static int x, y;
   int z;
   extern int* getPtr();
   int* p = &x;

   void callMe(int* q) {
       int w;
       int* r = getPtr();
       if (r == NULL)
           r = &w;
   }
\end{lstlisting}
}
\caption{Example of an incomplete program with pointers \texttt{p}, \texttt{q}, and \texttt{r}, all of which may point to unknown targets from external modules.}
\label{lst:unknown}
\end{figure}

The problem is illustrated in \autoref{lst:unknown}.
In this incomplete program, the pointers \texttt{q} and \texttt{r} originate from outside the current module, making it impossible for an analysis to determine the origins of the pointer values.
Nevertheless, it is possible to infer some facts about their targets.
Our analysis correctly infers that \texttt{p}, \texttt{q} and \texttt{r}
may target \texttt{x}, \texttt{z}, or any memory object defined in external modules, but never \texttt{y}.
Only \texttt{r} may target~\texttt{w}.

In contrast, other analyses require the presence of summary functions to render the program in \autoref{lst:unknown} complete~\cite{Pearce:SCAM03, Steensgaard:POPL1996, Hardekopf:PLDI2007}, or otherwise produce an unsound solution~\cite{Lei:SAS2019, Barbar:CGO2021, Barbar:OOPSLA2021}.
While unsound analyses might be acceptable for some tools or clients, such as in bug detection or refactoring assistance, optimizing compilers can not tolerate unsoundness~\cite{soundiness:CACM2015}.
Summary functions are not always available, such as when using external libraries, and can at most be an optional tool for improving the precision of common library functions.

In this paper, we present an Andersen-style points-to analysis~\cite{Andersen:dissertation1994} that produces sound solutions for all well-formed incomplete C~programs adhering to the standard provenance-aware memory model~\cite{ISO:Provenance6010}.
Our key insight is that a sound solution can be accomplished by tracking which memory locations and pointers are accessible to external modules.
Pointers with external origins may only target these externally accessible memory locations, avoiding the need for an overly conservative ``Unknown''-flag seen in other analyses~\cite{Andersen:dissertation1994, Lattner:PLDI2007}.
Importantly, a pointer value of unknown origin may not target any memory locations from the current module that have not escaped.
Our efficient implementation relies on six new constraint types, which enable us to implicitly represent externally accessible memory locations in the constraint graph, significantly reducing solver runtime.
We also present the \textit{Prefer Implicit Pointees} (PIP) technique that further reduces the use of explicit pointees by detecting situations where explicit pointees do not affect the final solution.

The results from running our analysis on a set of \numNonEmptyC C~files from nine SPEC CPU2017~\cite{SPEC-CPU2017} benchmarks and four open-source programs show that the implicit representation of pointees is the most important factor for efficient solving.
We achieve a total speedup of 15$\times$ over an oracle that always chooses the fastest configuration with explicit representation.
Enabling PIP in addition to representing pointees implicitly gives an additional 1.9$\times$ speedup.
The evaluation includes five additional solver techniques from the literature~\cite{Rountev:PLDI2000, Pearce:SCAM03, Hardekopf:PLDI2007}. In the context of individual file analysis, none of them improve upon PIP on average.
Finally, the precision of the analysis results is evaluated in terms of an alias-analysis client.
Compared to using a local IR-traversing alias analysis, LLVM's BasicAA, incorporating information from the points-to graph reduces the number of \texttt{MayAlias}-responses by 40\% on average.
The results show that our sound points-to analysis can provide additional alias information at scale, making it practical for production compilers.
\vspace{4mm}

\section{Background}
\label{sec:background}

A points-to analysis takes all pointers and memory objects in a program, and produces a points-to set $\Sol(p)$ for each pointer $p$.
The solution is \emph{sound} if $\Sol(p)$ contains all memory objects $p$ may ever target at runtime in all possible executions of the program.
$\Sol(p)$ may also contain \emph{spurious} memory objects that $p$ never targets,
which make the solution less \emph{precise}~\cite{Pearce:SQJ2004}.
An efficient analysis typically requires trading precision for scalability, and multiple trade-offs exist.
A \emph{flow-insensitive} analysis ignores control flow, and treats the program as an unordered set of statements~\cite{Pearce:SCAM03}.
A \emph{context-insensitive} analysis unifies arguments and return values between all calls to a function~\cite{Hind:PASTE2001}.
A \emph{field-insensitive} analysis represents all fields of an aggregate memory object using a single points-to set~\cite{Pearce:TOPLAS2007}.

Points-to analyses can typically be categorized as either \emph{unification-based} or \emph{inclusion-based}~\cite{Hind:PASTE2001},
where the latter models the flow of points-to sets directionally, giving more precise solutions.
Analyses that are inclusion-based and flow-insensitive are commonly referred to as \emph{Andersen-style}~\cite{Hind:PASTE2001},
and are the topic of this paper. We consider a field- and context-insensitive implementation.

\subsection{Building Constraint Sets}
\label{sec:building-constraint-sets}

In the first phase of an Andersen-style analysis, the program is converted into a finite set of abstract memory locations $M$, pointers $P$, and constraints $C$.
The constraints model all possible ways in which pointers may be given pointees, such as taking the address of memory objects, copying pointer values, loading and storing pointers, and passing pointers into and out of function calls.
We define a constraint language to represent these constraints, shown in \autoref{tab:constraints}.
The first four constraints are identical to the ones used by Pearce~\cite{Pearce:SCAM03}.
Function calls are handled similarly to Foster's $lam$ construct~\cite{Foster:Flow1997}, and support indirect function calls.

\begin{table}
    \centering
    \caption{Constraint types in an Andersen-style constraint language, and examples of C statements they represent.}%
    \label{tab:constraints}
    \resizebox{\columnwidth}{!}{
    \begin{tabular}{lll}
    \toprule
        Name & Constraint & Corresponding C code \\
    \midrule
        Base & $p \supseteq \{a\}$ & \texttt{p = \&a;} \\[1mm]
        Simple & $p \supseteq q$ & \texttt{p = q;} \\[1mm]
        Load & $p \supseteq {*q}$ & \texttt{p = *q;} \\[1mm]
        Store & ${*p} \supseteq q$ & \texttt{*p = q;} \\[1mm]
        Function & $\Func(f, r, a_1, ..., a_n)$ &
        \begin{tabular}{@{}l@{}}
        \texttt{void* f(a$_1$, $\ldots$, a$_n$)} \\
        \texttt{\{ $\ldots$; return r; \}}
        \end{tabular}
        \\[3mm]
        Function call & $\CallFunc(f, r, a_1, ..., a_n)$ & \texttt{r = (*f)(a$_1$, $\ldots$, a$_n$)} \\[1mm]
    \bottomrule
    \end{tabular}
    }
\end{table}

The exact process of converting the source program into constraints depends on the program representation.
Modern intermediate representations, like LLVM IR, make a distinction between variables stored in memory and temporary values stored in \emph{virtual registers}~\cite{llvm:CGO2004}.
Registers can not be pointed to, so the analysis only needs to consider them if their type is \emph{pointer compatible}.
Types that are not pointer compatible do not have points-to sets, and can be ignored by the analysis~\cite{Lattner:PLDI2007}.
In this work, we consider a type to be pointer-compatible if it is a pointer or an aggregate type containing a pointer.
The set of pointers $P$ includes all pointer-compatible virtual registers.

During runtime, the program may allocate arbitrarily many memory objects, which must be represented by a finite set of abstract memory locations $M$.
Named memory objects, such as local and global variables, functions, and imported symbols, are represented by distinct abstract memory locations.
Heap allocations are named after their allocation site, using distinct abstract memory locations to represent the memory objects allocated at each site.
Thus, each memory object in the program is represented by one, not necessarily unique, abstract memory location.
If an abstract memory location may represent values of pointer-compatible types, it is included in both $P$ and $M$.
The universe of constraint variables is denoted $V = P \cup M$.
The $\Func(\ldots)$ and $\CallFunc(\ldots)$ constraints ignore return values and arguments that are not pointer compatible.

\subsection{Solving Constraints}
\label{sec:solving-constraints}

The second analysis phase takes the sets $M$, $P$, and $C$ and solves the constraints to produce the final solution $\Sol:P \rightarrow \PowerSet(M)$.
To formalize how the constraints propagate points-to sets, a system of inference rules is used. The rules are shown in \autoref{fig:inference}, and mostly correspond to those of Pearce \etal~\cite{Pearce:TOPLAS2007}.

\begin{figure}
\begin{tabular}{@{}cccc@{}}
\textsc{trans} & \textsc{load} & \textsc{store} & \textsc{call} \\
& & & $h \supseteq \{f\}$ \\
$q \supseteq \{x\}$ & $q \supseteq \{x\}$ & $p \supseteq \{x\}$ & $\Func(f, r_{\bullet}, a_{1\bullet}, \cdots, a_{n\bullet})$ \\ 
$p \supseteq q$ & $p \supseteq {*q}$ & ${*p} \supseteq q$ & $\CallFunc(h, r, a_1, \cdots, a_n)$ \\
\cmidrule(r{1mm}){1-1} \cmidrule(lr{1mm}){2-2} \cmidrule(lr{1mm}){3-3} \cmidrule(l{1mm}){4-4}
$p \supseteq \{x\}$ & $p \supseteq x$ & $x \supseteq q$ & $r \supseteq r_{\bullet}$ \\
& & & $a_{i\bullet} \supseteq a_i, \forall i \in \{1, \ldots, n\}$ \\
\end{tabular}
\caption{Rules of inference for sound points-to set tracking}%
\label{fig:inference}
\end{figure}

\begin{figure}[t]
\begin{minipage}[t]{\linewidth}
\hspace{2cm} $P = \{p, q, r, s, x\}$ \hfill $M=\{x,y\}$ \hspace{2cm}
\end{minipage}
\smallskip
\begin{minipage}[b]{0.3\linewidth}
\begin{align*}
p &\supseteq \{x\}\\[0mm] 
q &\supseteq p \\[0mm]
r &\supseteq \{y\} \\[0mm]
{*q} &\supseteq r \\[0mm]
s &\supseteq {*q}
\end{align*}
\end{minipage}
\hspace{0.05\linewidth}
\begin{minipage}[b]{0.6\linewidth}
\includegraphics[width=0.95\textwidth]{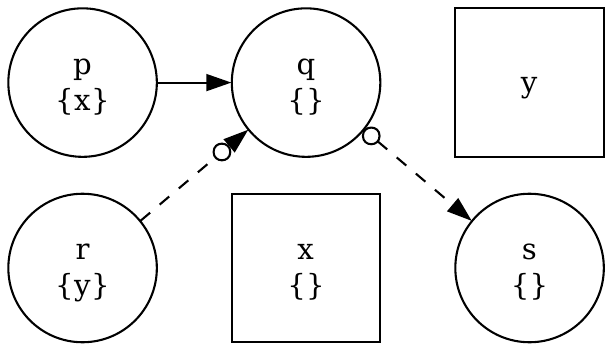}
\end{minipage}
\caption{A set of example constraints, and the corresponding constraint graph. Note that $y \notin P$, so there is no $\Sol(y)$.}
\label{fig:constraint-graph-simple}
\end{figure}

\noindent
The rules are designed to have the following property: "If it is possible for a pointer $p$ to point to a memory location $x$ at runtime, it is also possible to infer the constraint $p \supseteq \{x\}$"~\cite{Pearce:TOPLAS2007}. The analysis solution can thus be defined as:
\[
\Sol(p) \coloneq \{x \in M\ |\ p \supseteq \{x\} \text{ can be inferred from } C\}.
\]
The set of constraints can be formulated as a \emph{constraint graph}, as presented by Heintze and Tardieu~\cite{Heintze:SIGNOTICES2001}.
The technique has been refined and visualized in several works~\cite{Pearce:SCAM03, Pearce:TOPLAS2007, Hardekopf:PLDI2007, Pereira:CGO2009}.
In the constraint graph, the variables in $V$ are drawn as nodes,
while constraints are drawn as edges and labels. Nodes representing virtual registers are drawn as circles,
while abstract memory locations are squares.
The node representing a pointer $p \in P$ includes a braced list containing the targets of all its base constraints ($p \supseteq \{x\}$).
This list shows the current approximation of $\Sol(p)$.
Simple constraints ($p \supseteq q$) are drawn as simple edges $q \simplearrow p$.
Load and store constraints are drawn as the complex edges $\loadarrow$ and $\storearrow$, respectively, where the circle indicates the side being dereferenced.
\autoref{fig:constraint-graph-simple} shows an example of constraint variables and constraints, and the corresponding constraint graph.

\begin{figure}[t]
\begin{minipage}[c]{0.3\linewidth}
\begin{align*}
q &\supseteq \{x\} \\[0mm] %
x &\supseteq r \\[0mm] %
s &\supseteq x \\[0mm] %
x &\supseteq \{y\} \\[0mm] %
s &\supseteq \{y\} %
\end{align*}
\end{minipage}
\hspace{0.05\linewidth}
\begin{minipage}[c]{0.6\linewidth}
\includegraphics[width=0.95\textwidth]{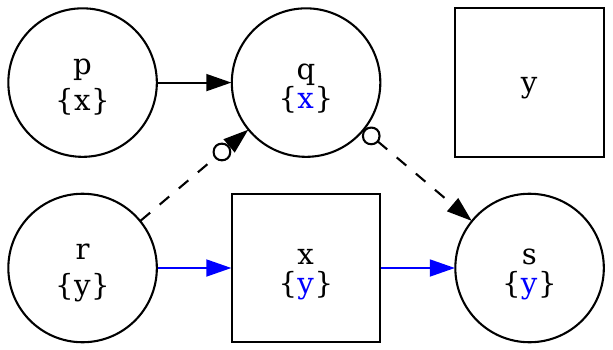}
\end{minipage}
\caption{The solved version of the constraint graph from \autoref{fig:constraint-graph-simple}. The inferred constraints are listed on the left, and colored blue in the graph.}
\label{fig:constraint-graph-simple-solved}
\end{figure}

The inference rules from \autoref{fig:inference} can all be represented as operations on the constraint graph,
using existing constraints to infer new ones.
The \textsc{trans} rule corresponds to propagating the contents of $\Sol$ sets along simple edges.
The \textsc{load} and \textsc{store} rules correspond to converting complex edges into simple edges by dereferencing the end with the circle.
\autoref{fig:constraint-graph-simple-solved} shows inferring all possible constraints starting with the set in \autoref{fig:constraint-graph-simple}, by using the constraint graph.

\begin{figure}
\begin{minipage}[c]{0.2\textwidth}
{\footnotesize
\begin{lstlisting}[language=C,texcl, numbers=left,numbersep=-6pt]
   int* f(int** x) {
     int* y = *x;
     return y;           
   }
   int*(*h)(int**) = &f;

   int a;
   int* g() {
     int* p = &a;
     int** r = &p;
     int* s = (*h)(r);
     return s;
   }
\end{lstlisting}
}
\end{minipage}
\begin{minipage}[c]{0.5\textwidth}
    \includegraphics[width=0.64\columnwidth]{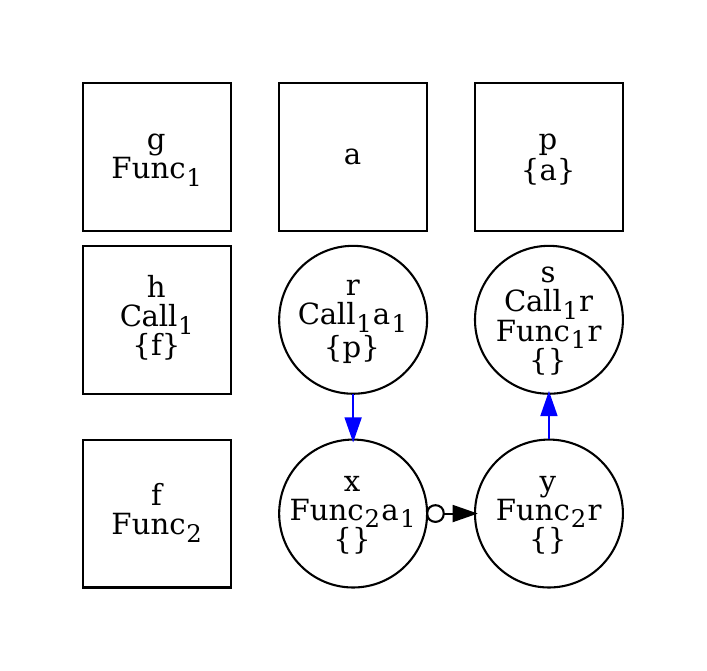}
\end{minipage}
\caption{A sample C program containing two functions and an indirect function call. The corresponding constraint graph is drawn in black.
The result of applying the \textsc{call} inference rule to $\CallFunc_1$ and $\Func_2$ is drawn in blue.
Local variables that never have their address taken are represented by virtual registers, drawn as circles. The remaining constraint variables are abstract memory locations, drawn as squares.}%
\label{fig:constraint-graph-functions}
\end{figure}

The constraint graph can be extended to include function calls
by assigning each $\Func$ and $\CallFunc$ constraint a unique number,
and adding labels to nodes representing functions, call targets, arguments, and return values.
The \textsc{call} inference rule can be applied when a node labeled $\CallFunc_i$ has a node labeled $\Func_j$ in its $\Sol$ set.
Applying the rule adds simple edges representing the flow of possible pointees through arguments
$\CallFunc_{i}a_{n} \rightarrow \Func_{j}a_{n}$ and through the return value $\Func_{j}r \rightarrow \CallFunc_{i}r$.
\autoref{fig:constraint-graph-functions} shows a C program with two functions and the corresponding constraint graph.

\subsection{Worklist Solver}
\label{sec:worklist}

Multiple variations of worklist algorithms have been used to solve constraint graphs efficiently~\cite{Pearce:SCAM03, Hardekopf:PLDI2007}. The worklist is a list of nodes that need to be visited before solving is finished.
Visiting a node checks the constraints currently defined on the node to see if any of the inference rules from \autoref{fig:inference} can be used to infer new constraints.
When a node receives new constraints, it is added to the worklist to ensure the new constraints are processed.
Once the worklist is empty, all constraints have been processed, and a fixed point has been reached.
The order in which nodes from the worklist are processed is known as its \emph{iteration order}, and can have a drastic effect on solving performance~\cite{Hardekopf:PLDI2007}.

\subsection{Cycle Detection}

An important observation made by Fähndrich \etal is that cycles in the constraint graph can be eliminated~\cite{Fahndrich:PLDI1998}.
When a cycle of simple edges $a \rightarrow b \rightarrow c \rightarrow \ldots \rightarrow a$ is formed, any member of $\Sol(a)$ also becomes a member of $\Sol(b)$, $\Sol(c)$, etc.
Instead of propagating pointees through the cycle, the pointers in the cycle can share a single $\Sol$ set, saving time and memory.
Several techniques exist for discovering such cycles, both before~\cite{Rountev:PLDI2000,Hardekopf:SAS2007} and during~\cite{Pearce:SCAM03,Hardekopf:PLDI2007,Pereira:CGO2009} solving.

\subsection{Pointer Provenance}

The C standard places restrictions on the use of pointers, prohibiting out-of-bounds pointer arithmetic or using pointers to freed memory~\cite{ISO:C23}.
This can lead to situations where two pointers have the same address, yet are non-interchangeable,
as demonstrated in Defect Report DR260.
The resolution from the C language working group WG14 confirmed:
``[Implementations may] treat pointers based on different origins as distinct even though they are bitwise identical''~\cite{Feather:DR260}.
The origin of a pointer is known as its \emph{provenance},
and compilers routinely exploit this fact when performing optimizations~\cite{Memarian:POPL2019}.

Some points-to analyses handle pointer-integer casting by tracking the pointees of integers~\cite{Steensgaard:POPL1996, Lattner:PLDI2007}.
This may have unintended side effects, however, as it adds the concept of provenance to integers.
Integers with identical representation may no longer be interchangeable,
which breaks common assumptions, and has led to miscompilations~\cite{Lee:OOPSLA2018}.
When provenance was standardized in Technical Specification 6010~\cite{ISO:Provenance6010}, the committee therefore went with
a provenance-aware memory object model called \emph{PNVI-ae-udi}~\cite{Gustedt:Provenance2021}.
In this model, provenance is not carried via integers (\emph{PNVI}).
Instead, casting an integer to a pointer recreates the provenance of the memory object it targets.
The target must, however, have previously had its address exposed (\emph{ae}) as an integer.

\section{Handling Incomplete Programs}
\label{sec:incomplete_programs}

The analysis from \autoref{sec:background} only produces sound solutions when analyzing whole programs.
This prevents its use in situations where only parts of the source code are available, such as when compiling a single file in a larger program or when using external libraries. In these situations, the analysis is said to operate on \emph{incomplete programs}~\cite{Hind:PASTE2001, Lattner:PLDI2007}.

In this section, we show how to extend the analysis to make it sound for incomplete programs.
The key insight is that this can be accomplished by tracking which memory locations escape the module, and which pointers originate from external modules.
Production C compilers also need to handle pointer-integer conversions, which are covered in \autoref{sec:integer_conversion}.
The resulting analysis is sound for all well-formed incomplete C programs, but its naive implementation is slow, as shown in \autoref{sec:results}.
A more efficient implementation is presented in \autoref{sec:implicit_pointees}.

\subsection{Tracking Externally Accessible Locations and Pointers}
\label{sec:tracking-externally-accessible-memory}

An incomplete program is always executed as part of a larger complete program, which may include arbitrary statements and variables from other modules.
We define the soundness of points-to analysis solutions for incomplete programs as follows:
\textit{%
For an incomplete program $A$ with pointers $P$ and abstract memory locations $M$, a points-to analysis solution $\Sol : P \rightarrow \PowerSet(M)$ is sound if and only if\ \ $\Sol(p)$ contains all pointees $p$ may ever target at runtime, during any execution of any complete program $A$ is linked into.
}

We call the incomplete program being analyzed the \emph{current module}, and all other modules \emph{external modules}.
To conservatively model all possible statements in external modules,
the analysis keeps track of which abstract memory locations in $M$ \emph{escape} the current module.
A memory location $x \in M$ may escape when:
\begin{itemize}
    \item $x$ is exported from the current module as a named symbol.
    \item A pointer to $x$ is an argument to an external function call.
    \item A pointer to $x$ is the return value in a function $f$, where $f$ itself has escaped and may be called by external modules.
    \item $x \in \Sol(y)$, and $y$ has escaped.
\end{itemize}
Memory locations that escape from the current module, and memory locations imported from other modules, make up the set of \emph{externally accessible} memory locations.
Tracking which memory locations are in this set lets the analysis retain some precision when handling pointer values of unknown origin.
The situations in which a pointer $p$ has an unknown origin are:
\begin{itemize}
    \item $p$ is the return value of an external function call.
    \item $p$ is an argument in an escaped function $f$ that may be called from an external module.
    \item $p$ is a memory location that is externally accessible. An external module may thus store a new pointer into~$p$.
\end{itemize}
These cases correspond to the pointers \texttt{r}, \texttt{q}, and \texttt{p} from \autoref{lst:unknown}, respectively.
In all these cases, the pointer values originate from external modules. This means they may only target externally accessible memory locations.
Importantly, a pointer value of unknown origin may not target any memory locations from the current module that have not escaped.
Even if an external module is able to correctly guess the address of a non-escaped memory object, pointers created this way would not have the correct provenance, and would be invalid~\cite{Memarian:POPL2019}.

\subsection{The \external Node}
\label{sec:external-node}

A new constraint variable is introduced, called \external, to keep track of externally accessible memory locations.
It represents all memory locations defined in external modules that are not represented by any other abstract memory location.
This memory may contain pointers, so $\external \in P$.
Effectively, $\Sol(\external)$ contains all memory locations that are externally accessible.
It can also be pointed to, so $\external \in M$.
Along with the \external node, the following constraints are added:
\[
\begin{array}{wl{23mm} wl{23mm} wl{23mm}}
    \circled{1}\ \ \external \supseteq \{\external\} &
    \circled{2}\ \ \external \supseteq {*\external} &
    \circled{3}\ {*\external} \supseteq \external
\end{array}
\]
\[
\begin{array}{wl{36mm} wl{36mm}}
    \circled{4}\ \ \CallFunc(\external, \external, \cdots, \external) &
    \circled{5}\ \ \Func(\external, \external, \cdots, \external)
\end{array}
\]
Constraint \circled{1} represents the fact that any pointer represented by \external may target memory represented by \external. This also gives the desired property that loading from an unknown pointer results in another unknown pointer.
Any pointer $p$ with an unknown origin is given the constraint $p \supseteq \external$, which makes $p$ target \external itself, as well as all externally accessible memory locations.

\begin{figure}
\centering
\includegraphics[width=.98\linewidth]{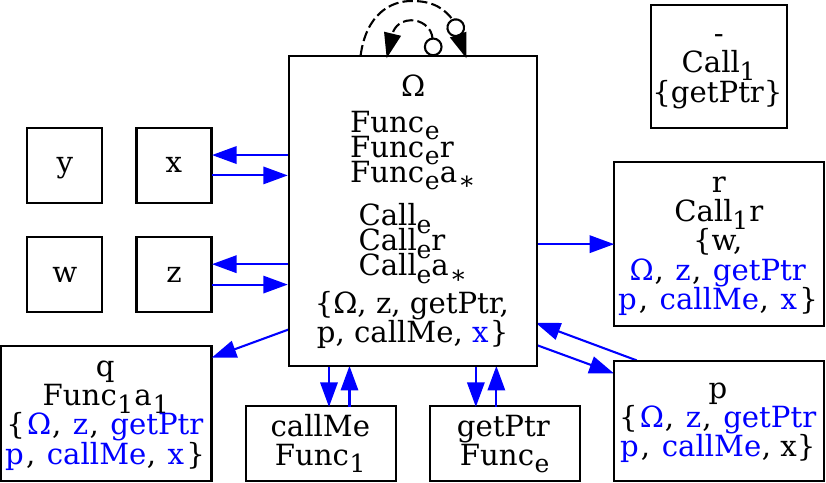}
\caption{The constraint graph corresponding to the program in \autoref{lst:unknown}. Black constraints are added during phase 1, and blue constraints are added during solving. The direct function call on line 10 has been given a dummy pointer to \texttt{getPtr} to be the target of the $\CallFunc_1$ constraint.}
\label{fig:example-with-external-node}
\end{figure}

The remaining constraints are used to conservatively represent all statements that may ever occur in any external module.
Constraint \circled{2} represents external modules loading from any pointer they may hold, causing the pointees of escaped pointers to also escape. Constraint \circled{3} represents external modules storing unknown pointer values into any memory that has escaped, giving all escaped pointers unknown pointees.

Constraint \circled{4} handles functions that escape from the current module being called from external modules. Pointer-compatible return values and function arguments are propagated to and from \external, respectively.
Constraint \circled{5} handles cases where indirect function calls target an unknown pointer, which may target functions in external modules.

Finally, copies of constraint \circled{5} also need to be added to all nodes representing functions imported from other modules. Recall that the \external node only represents external memory that is not otherwise represented by any other abstract memory location, so imported functions need their own $\Func(\cdot)$ constraint.
If the imported function is a common library function, it is also possible to use a handwritten summary function instead of the overly conservative constraint \circled{5}.

The constraints on the \external node effectively model all statements that may be executed in external modules.
The constraints are expressed using the regular constraint language, which means existing constraint graph solvers can support the \external node with only minor modifications.
The main modifications required to handle incomplete programs are the following additions to the first phase of the analysis:
\begin{itemize}
\item The \external node and its constraints are added.
\item Any exported global variable or function $e$ is marked as externally accessible ($\external \supseteq \{e\}$).
\item Any imported global variable or function $i$ is marked as externally accessible ($\external \supseteq \{i\}$).
\item Any imported function $f$ must either be represented by a generic constraint $\Func(f, \external, \cdots, \external)$, or be mapped to a custom summary function for $f$.
\end{itemize}
The additions to the constraint graph only consider top-level definitions and declarations from the analyzed program. This is sufficient, as named symbols form the basis of all cross-module interactions.
The constraints on the \external node ensure that all statements that may be executed in external modules are reflected in the current module's analysis solution.
\autoref{fig:example-with-external-node} shows the constraint graph for the program in \autoref{lst:unknown}.

In the example program, the exported symbols are \texttt{z}, \texttt{p}, and \texttt{callMe}, while the only imported symbol is \texttt{getPtr}. These are all externally accessible as symbols, and are thus included in the initial $\Sol(\external)$ set. Since \texttt{getPtr} is an externally defined function, it is also given the $\Func_e$ constraint, whose return value and arguments are all represented by \external.

The constraints defined on the external node ensure that all statements that may be executed in external modules are handled when the graph is solved. The \external load and store self-edges add simple edges to and from all memory locations in $\Sol(\external)$. One of these edges, $p \simplearrow \external$, causes $x$ to be propagated from $\Sol(p)$ to $\Sol(\external)$, representing $x$'s escape from the current module. The edge $\external \simplearrow p$ adds all externally accessible memory locations to $\Sol(p)$, representing the fact that external modules may store pointer values of unknown origin into $p$.

The $\CallFunc_e$ constraint on \external represents external modules calling all externally accessible functions, such as \texttt{callMe}. Applying the \textsc{call} inference rule adds the simple edge $\external \simplearrow q$. Every externally accessible memory location is propagated along this edge, representing that $q$ may hold a pointer value of unknown origin.
Likewise, the function call on line 10 targets \texttt{getPtr}, which has the $\Func_e$ constraint, adding the simple edge $\external \simplearrow r$ to represent pointer values of unknown origin. $r$ itself does not escape, so it can also point to $w$ without making $w$ externally accessible.

\begin{figure*}
\centering
\begin{tabular}{ccccccc}
\textsc{In\external} & \textsc{Trans\external} & \textsc{To\external} & \textsc{StoreTo\external} & \textsc{LoadFrom\external} & \textsc{StoreScalar} & \textsc{LoadScalar} \\

\inference{\external \sqsupseteq \{p\}}{\begin{array}{c}
\external \sqsupseteq p \\
p \sqsupseteq \external
\end{array}} &

\inference{p \sqsupseteq \external \\ 
q \supseteq p}{q \sqsupseteq \external} &

\inference{\external \sqsupseteq p \\ p \supseteq \{x\}}{
\external \sqsupseteq \{x\}
} &

\inference{{*p} \supseteq q \\ p \sqsupseteq \external }{
\external \sqsupseteq q
} &

\inference{q \supseteq {*p} \\ p \sqsupseteq \external}{
q \sqsupseteq \external
} &

\inference{{*p} \sqsupseteq \external \\ p \supseteq \{x\}}{
x \sqsupseteq \external
} &

\inference{\external \sqsupseteq {*p} \\ p \supseteq \{x\}}{
\external \sqsupseteq x} \\
\end{tabular}

\begin{tabular}{ccccc}
\textsc{CallImp} & \textsc{Call\external} & \textsc{CalledBy\external} \\

\inference{
\CallFunc(f,r,a_1,\cdots,a_k) \\
f \supseteq \{g\} \\
\ImportedFunc(g)
}{\begin{array}{c}
r \sqsupseteq \external\\
\external \sqsupseteq a_i,\ \forall i \in \{1, \cdots, k\}
\end{array}} &

\inference{
\CallFunc(f,r,a_1,\cdots,a_k) \\
f \sqsupseteq \external
}{\begin{array}{c}
r \sqsupseteq \external\\
\external \sqsupseteq a_i,\ \forall i \in \{1, \cdots, k\}
\end{array}} &

\inference{
\Func(f,r,a_1,\cdots,a_k) \\
\external \sqsupseteq \{f\}
}{\begin{array}{c}
\external \sqsupseteq r\\
a_i \sqsupseteq \external,\ \forall i \in \{1, \cdots, k\}
\end{array}} \\
\end{tabular}
\caption{Additional rules of inference added to represent the \external node implicitly.}%
\label{fig:implicit-inference}
\end{figure*}

\subsection{Pointer-Integer Conversions and Pointer Smuggling}
\label{sec:integer_conversion}

To make the points-to analysis sound under the PNVI-ae-udi provenance model, integers can not be pointer compatible.
Instead, when a pointer $p$ is cast to an integer, all pointees of $p$ are marked as externally accessible by adding the constraint $\external \supseteq p$. When an integer is cast to a pointer $q$, it soundly handles having an unknown origin by adding the constraint $q \supseteq \external$. The flow of pointer values via integers is thus soundly represented via the \external node. This also handles cases where pointers are cast to integers in the current module, and cast back to pointers in an external module, or vice versa.

Since integers are not pointer compatible, a memory location $x$ representing an integer variable is not included in the set of pointers $P$.
The analysis does not track the pointees of $x$, and there is no $\Sol(x)$ set. It is, however, still possible for constraints of the form $p \supseteq x$ or $x \supseteq p$ to appear, where $p \in P$. These constraints effectively represent conversions between pointers and integers, so a sound analysis must instead treat these constraints as $p \supseteq \external$ or $\external \supseteq p$, respectively.

The last piece necessary to make the analysis sound is to handle indirect casting of pointer types through type punning.
By casting an \texttt{int**} to a \texttt{char*}, and reading eight consecutive \texttt{char}s, an \texttt{int*} has effectively been converted from a pointer to a scalar.
Reversing this process converts the scalar back into a pointer.
We call this indirect casting via memory \emph{pointer smuggling}.

Pointer smuggling can be handled soundly by adding constraints on loads and stores of pointer-incompatible types. Given a \texttt{char*} $p$, loading a \texttt{char} from $p$ adds the constraint $\external \supseteq {*p}$, and storing a \texttt{char} to $p$ adds the constraint ${*p} \supseteq \external$. If $p$ happens to have a pointer-compatible target $q \in \Sol(p)$, it will correctly be marked as $\external \supseteq q$ and $q \supseteq \external$, respectively.

\subsection{Representing \external implicitly (\implicit)}
\label{sec:implicit_pointees}

The introduction of the \external node enables the sound points-to analysis of incomplete programs, expressed within the conventional language of constraints for Andersen-style points-to analysis. The \external node is, however, likely to be a hot spot during solving. 
Every externally accessible memory location is included in $\Sol(\external)$, and all pointers of unknown origin have $\Sol$ sets that are supersets of $\Sol(\external)$.
This Cartesian product of pointer-pointee relations scales poorly, both in memory usage and solver runtime.
Cycle detection is of limited help, as can be seen in the solved constraint graph in \autoref{fig:example-with-external-node}. Only $p$ is in a cycle with \external, while $q$ and $r$ are not.

The key observation to make the analysis scalable is that the \external node can be represented entirely implicitly. The \external node is removed, and six new types of constraints are added to the constraint language in its place, replacing constraints previously defined using the \external node. The new constraint types, shown in \autoref{tab:ip_constraints}, use the symbol $\sqsupseteq$ to clearly distinguish them from the original constraint language. In the new language, \external is strictly a language construct, not a constraint variable.

Turning \external into a part of the language allows us to implement the behavior described in \autoref{sec:external-node}
using an alternative set of inference rules, shown in \autoref{fig:implicit-inference}.
The new rule \textsc{Trans\external} avoids propagating copies of $\Sol(\external)$ around the constraint graph,
by instead propagating the constraint $p \sqsupseteq \external$ itself along simple edges $p \simplearrow q$.
The rule \textsc{To\external} marks targets sent to \external as externally accessible, while \textsc{In\external} emulates the store and load self-edges of the \external node.
The \textsc{StoreTo\external} and \textsc{LoadFrom\external} rules handle storing and loading of pointers that may have an unknown origin, while
the \textsc{Call*} rules emulate the $\Func$ and $\CallFunc$ constraints on \external.
In total, the new rules replicate all behavior that was previously modeled by the \external node, without inferring any constraints from the original language.

\begin{table}[t]
    \centering
    \caption{Constraint types added in the extended language to represent the \external node implicitly.}%
    \label{tab:ip_constraints}
\resizebox{\columnwidth}{!}{
\setlength{\tabcolsep}{1mm}
\begin{tabular}{lll}
\toprule
Old & New & Comment \\
\midrule
$\external \supseteq \{x\}$ & $\external \sqsupseteq \{x\}$ & $x$ is externally accessible \\
$p \supseteq \external$ & $p \sqsupseteq \external$ & $p$ targets all externally accessible memory \\
$\external \supseteq p$ & $\external \sqsupseteq p$ & Pointees of $p$ are externally accessible \\
${*p} \supseteq \external$ & ${*p} \sqsupseteq \external$ & A scalar is stored at $*p$ \\
$\external \supseteq {*p}$ & $\external \sqsupseteq {*p}$ & $*p$ is loaded as a scalar \\
$\Func(f, \external, \cdots, \external)$ & $\ImportedFunc(f)$ & $f$ is an imported external function \\
\bottomrule
\end{tabular}
}
\end{table}

There are now two ways of encoding that a pointer $p$ may point to a target memory location $x$. If the solver infers the base constraint $p \supseteq \{x\}$, we say that $x$ is an \emph{explicit} pointee of $p$.
If the solver infers both constraints $p \sqsupseteq \external$ and $\external \sqsupseteq \{x\}$, we say that $x$ is an \emph{implicit} pointee of $p$.
While it would be possible to add a rule that infers explicit pointees from implicit pointees, this rule has been omitted on purpose. Instead, the definition of $\Sol$ is changed to include both explicitly and implicitly encoded pointers, defined as $\Sol_e$ and $\Sol_i$, respectively:
\begin{align*}
\Sol_e(p) &\coloneq \{x \in M\ |\ p \supseteq \{x\} \text{ has been inferred}\} \\
\Sol_i(p) &\coloneq \begin{cases}
E, & p \sqsupseteq \external \text{ has been inferred} \\
\{\},& \text{otherwise}
\end{cases} \\
\Sol(p) &\coloneq \Sol_e(p) \cup \Sol_i(p)
\end{align*}
\[
\text{where } E = \{x \in M\ |\ \external \sqsupseteq \{x\} \text{ has been inferred}\}
\]
The final $\Sol$ sets end up being identical to the solution produced using the explicit \external node. The benefit of introducing the implicit pointee representation is that many pointer-pointee relations only exist because all pointers of unknown origin may target all externally accessible memory locations.
Representing these possible pointer-pointee pairs explicitly would require one base constraint per pair, which scales poorly both in memory usage and solver runtime.
The implicit representation keeps the $\external \sqsupseteq \{x\}$ and $p \sqsupseteq \external$ constraints separate, and leaves the Cartesian product of pointer-pointee relations implicit.
Only pointer-pointee relations that go via \external can be represented implicitly, so explicit pointees are still used to represent all other kinds of pointer-pointee relations.
The conventional worklist solver can also be modified to include the new inference rules shown in \autoref{fig:implicit-inference}. The new constraint types can all be implemented as 1-bit flags on constraint variables. Pseudocode for the modified worklist algorithm is given in \autoref{alg:worklist}. 

\begin{algorithm}
  \caption{Worklist algorithm for inferring all constraints from \autoref{fig:inference} and \autoref{fig:implicit-inference}. The blue comments show where to add extra logic for the PIP technique.}%
  \label{alg:worklist}
  \begin{algorithmic}
    \small
    \Require{Sets of pointers $P$, memory locations $M$ and constraints $C$}
    \State $\Sol_e(p) \gets \{x \in M~|~(p \supseteq \{x\}) \in C\},\ \forall p \in P$ \Comment{Initialize $\Sol_e$}
        \Procedure{PropagatePointees}{$f,t$} \Comment{Propagates from $f$ to $t$}
    \State {$\Sol_e(t) \gets \Sol_e(t) \cup \Sol_e(f)$}
    \If{$f$ \textbf{is marked} $f \sqsupseteq \external$}
    \textbf{mark} $t$ \textbf{as} $t \sqsupseteq \external$
    \EndIf
    \If{$\Sol_e(t)$ \textbf{changed or} $t$ \textbf{was marked}}
    $W \gets W \cup \{t\}$
    \EndIf
    \EndProcedure
    \Procedure{CallToImported}{$r, a_1, \cdots, a_k$}
    \State{\textbf{mark} $r$ \textbf{as} $r \sqsupseteq \external$}
    \State{\textbf{mark} $a_i$ \textbf{as} $\external \sqsupseteq a_i,\ \ \forall i\in\{1, \cdots, k\}$}
    \State{$W \gets W \cup \{v \in (P \cup M)\ |\ v \text{ gained a flag in this procedure}\}$}
    \EndProcedure
    \Procedure{MarkExternallyAccessible}{$x$}
    \State{\textbf{mark} $x$ \textbf{as} $\external \sqsupseteq \{x\}$ \textbf{and} $x \sqsupseteq \external$ \textbf{and} $\external \sqsupseteq x$}
    \ForAll{$\Func(x, r, a_1, \cdots, a_k) \in C$}
    \State{\textbf{mark} $r$ \textbf{as} $\external \sqsupseteq r$}
    \State{\textbf{mark} $a_i$ \textbf{as} $a_i \sqsupseteq \external,\ \ \forall i\in \{1,\cdots,k\}$}
    \EndFor
    \State{$W \gets W \cup \{v \in (P \cup M)\ |\ v \text{ gained a flag in this procedure}\}$}
    \EndProcedure
    \LComment{Handle nodes that are externally accessible from the start}
    \ForAll{$x$ \textbf{marked} $\external \sqsupseteq \{x\}$}
    \State{\Call{MarkExternallyAccessible}{$x$}}
    \EndFor
    \LComment{Initialize Worklist with every node}
    \State $W \gets P \cup M$
    \While{$W \neq \emptyset$}
    \State{$n \gets \Call{PopWorklist}{W}$}
    \State{$\Delta E \gets \{\}$} \Comment{Simple edges to add}
    %
    %
    \LComment{\textcolor{blue}{PIP addition 1: Backpropagate to make $\external \sqsupseteq n$ if possible}}
    \If{$n$ \textbf{is marked} $\external \sqsupseteq n$}
    \ForAll{$x \in Sol_e(n)$}
    \If{$x$ \textbf{not marked} $\external \sqsupseteq \{x\}$}
    \State{\Call{MarkExternallyAccessible}{$x$}}
    \EndIf
    \EndFor
    \EndIf
    \LComment{\textcolor{blue}{PIP addition 2: If $\external \sqsupseteq n$ and $n \sqsupseteq \external$, clear $\Sol_e(n)$}}
    
    \ForAll{$p \supseteq n$ \textbf{in} $C$} \Comment{Simple edges}
    \LComment{\textcolor{blue}{PIP addition 4: If $p \sqsupseteq \external$ and $\external \sqsupseteq n$, remove the edge}}
    \State{\Call{PropagatePointees}{$n, p$}}
    \EndFor
    \ForAll{${*n} \supseteq p$ \textbf{in} $C$} \Comment{Store edges}
    \ForAll{$x \in \Sol_e(n)$}
    \State{$\Delta E \gets \Delta E \cup \{x \supseteq p\}$}
    \EndFor
    \If{$n$ \textbf{is marked} $n \sqsupseteq \external$}
    \State{\textbf{mark} $p$ \textbf{as} $\external \sqsupseteq p$, \textbf{add} $p$ \textbf{to} $W$ \textbf{if mark is new}}
    \EndIf
    \EndFor
    %
    \If{$n$ \textbf{is marked} ${*n} \sqsupseteq \external$} \Comment{Storing scalar}
    \ForAll{$x \in \Sol_e(n)$}
    \State{\textbf{mark} $x$ \textbf{as} $x \sqsupseteq \external$, \textbf{add} $x$ \textbf{to} $W$ \textbf{if mark is new}}
    \EndFor
    \EndIf
    %
    \ForAll{$p \supseteq {*n}$ \textbf{in} $C$} \Comment{Load edges}
    \ForAll{$x \in \Sol_e(n)$}
    \State{$\Delta E \gets \Delta E \cup \{p \supseteq x\}$}
    \EndFor
    \If{$n$ \textbf{is marked} $n \sqsupseteq \external$}
    \State{\textbf{mark} $p$ \textbf{as} $p \sqsupseteq \external$, \textbf{add} $p$ \textbf{to} $W$ \textbf{if mark is new}}
    \EndIf
    \EndFor
    %
    \If{$n$ \textbf{is marked} $\external \sqsupseteq {*n}$} \Comment{Loading scalar}
    \ForAll{$x \in \Sol_e(n)$}
    \State{\textbf{mark} $x$ \textbf{as} $\external \sqsupseteq x$, \textbf{add} $x$ \textbf{to} $W$ \textbf{if mark is new}}
    \EndFor
    \EndIf
    %
    \ForAll{$\CallFunc(n, r, a_1, \cdots, a_k)$ \textbf{in} $C$} \Comment{Calls}
    \ForAll{$x \in \Sol_e(n)$}
    \ForAll{$\Func(x, r_{\bullet}, a_{1\bullet}, \cdots, a_{k\bullet})$ \textbf{in} $C$}
    \State{$\Delta E \gets \Delta E \cup \{r \supseteq r_\bullet, a_{1\bullet} \supseteq a_1, \cdots, a_{k\bullet} \supseteq a_k\}$}
    \EndFor
    \If{$x$ \textbf{is marked} $\ImportedFunc(x)$}
    \State{\Call{CallToImported}{$r, a_1, \cdots, a_k$}}
    \EndIf
    \EndFor
    \If {$n$ \textbf{is marked} $n \sqsupseteq \external$} 
    \State{\Call{CallToImported}{$r, a_1, \cdots, a_k$}}
    \EndIf
    \EndFor
    \ForAll{$(p \supseteq q) \in \Delta E \backslash C$} \Comment{Add new simple edges}
    \LComment{\textcolor{blue}{PIP addition 3: Skip adding edge if $p \sqsupseteq \external$ and $\external \sqsupseteq q$}}
    \State{$C \gets C \cup \{p \supseteq q\}$}
    \State{\Call{PropagatePointees}{$q, p$}}
    \EndFor
    \EndWhile
  \end{algorithmic}
\end{algorithm}

\section{Prefer Implicit Pointees (PIP)}
\label{sec:pip}
The introduction of the implicit pointee representation gives the analysis an efficient way to encode the pointer-pointee relations between all pointers of unknown origin and all externally accessible memory locations.
It is, however, possible for a pointee to be represented both implicitly and explicitly, \ie, $x \in \Sol_e(p) \land x \in \Sol_i(p)$. This happens when the solver infers all three of the following constraints:
\[
p \supseteq \{x\} \hspace{1.5cm} p \sqsupseteq \external \hspace{1.5cm} \external \sqsupseteq \{x\}
\]
In these cases, we say that $x$ is a \emph{doubled-up} pointee of $p$.
These doubled-up pointees do not affect the final solution, but are undesired, as they have a higher representational overhead. Explicit pointees may also be propagated along simple edges or be used to infer additional simple edges, which can only serve to create more doubled-up pointees. This extra work increases runtime, without ever affecting the analysis solution.

Prefer Implicit Pointees (PIP) is an online solver technique that attempts to reduce the number of doubled-up pointees that occur during solving. It is based on the observation that for any node $p$ with the constraints $\external \sqsupseteq p$ and $p \sqsupseteq \external$, the final solution will always satisfy $\Sol(p) = \Sol_i(p)$. Thus, any explicit pointee of $p$ is guaranteed to end up as a doubled-up pointee, and can safely be removed, or never added in the first place.
In practice, the technique works by adding the following checks to the worklist algorithm presented in \autoref{alg:worklist}:
\begin{enumerate}[leftmargin=*]
\item When the algorithm visits a node $p$, and $p$ is not marked $\external \sqsupseteq p$, it starts by checking if $p$ can gain the flag through ``backpropagation''. This is possible if there exists an outgoing simple edge $p \rightarrow q$, where $q$ is marked $\external \sqsupseteq q$. In that case, any member $x \in \Sol_e(p)$ will be propagated to $\Sol_e(q)$, where it will be marked $\external \sqsupseteq \{x\}$. Adding $\external \sqsupseteq p$ thus does not affect the final solution.
\item If a node $p$ is marked both $p \sqsupseteq \external$ and $\external \sqsupseteq p$, the set $\Sol_e(p)$ can be cleared, after all members $x \in \Sol_e(p)$ have been marked $\external \sqsupseteq \{x\}$.
\item When attempting to add an edge $p \rightarrow q$, first check if the node $q$ has the constraint $\external \sqsupseteq q$. If so, backpropagate it to $\external \sqsupseteq p$. If $\external \sqsupseteq p$, and $q \sqsupseteq \external$, the addition of the simple edge $p \rightarrow q$ can be skipped entirely, as $\Sol(q)$ becomes a superset of $\Sol(p)$ via \external.
\item Before propagating explicit pointees along a simple edge $p \rightarrow q$, check if the nodes are marked $q \sqsupseteq \external$ and $\external \sqsupseteq p$. If so, the simple edge can be removed, using the same reasoning as in the previous point.
\end{enumerate}
By adding PIP, the worklist algorithm gains some important properties: Any node $x$ that is marked both $x \sqsupseteq \external$ and $\external \sqsupseteq x$ will be visited by the worklist at most once. $\Sol_e(x)$ will also be empty in the final solution.
Nodes like $x$ are common, as they include all externally accessible memory locations.

\section{Methodology}
\label{sec:methodology}

The described analysis has been implemented in the \texttt{jlm} research compiler~\cite{jlm:GITHUB2024}, which converts LLVM IR~\cite{llvm:CGO2004} into an intermediate representation based on the Regionalized Value State Dependence Graph (RVSDG)~\cite{rvsdg:TECS2020}. All instructions that are relevant to points-to analysis in the original LLVM IR have a one-to-one representation in the RVSDG, making the results applicable to LLVM IR or other SSA-based IRs.
The analysis is performed on all nine C benchmarks from SPEC2017~\cite{SPEC-CPU2017}, and four C programs often used in the literature~\cite{Pearce:TOPLAS2007,Pereira:CGO2009,Nasre:CGO2011}.
The benchmarks are summarized in \autoref{tab:benchmarks}, totaling \numNonEmptyC non-empty C files.
Evaluation is performed in the context of a conventional compilation model,
where each C file is analyzed separately.
The files are converted to LLVM IR by \texttt{clang 18.1.8} using optimization level \texttt{O0},
and analyzed by \texttt{jlm} on an Intel Xeon Gold 6348 CPU running at 3.0 GHz.

\begin{table}
    \caption{Programs used to benchmark points-to analysis runtime and precision. $V$ is the set of constraint variables, and $C$ is the set of constraints, per analyzed file.}%
    \label{tab:benchmarks}
    \setlength{\tabcolsep}{3.5pt}
    \resizebox{\columnwidth}{!}{
    \begin{tabular}{lrrrrrrrr}
    \toprule
 & & \multicolumn{7}{c}{Summary of non-empty C files} \\
 \cmidrule(lr){3-9}
 & & & \multicolumn{2}{c}{IR instructions} & \multicolumn{2}{c}{$|V|$} & \multicolumn{2}{c}{$|C|$} \\
 \cmidrule(lr){4-5} \cmidrule(lr){6-7} \cmidrule(lr){8-9}
\multicolumn{2}{c}{Name \hfill KLOC$^\dagger$} & \#Files & Mean & Max & Mean & Max & Mean & Max \\
\midrule
500.perlbench & 362 &      68 &  22\;725 & 165\;497 &   4\;226 &  28\;236 &   6\;686 &  47\;046 \\
502.gcc       & 902 &      372 &  16\;244 & 535\;524 &   3\;434 &  64\;847 &   5\;148 & 101\;572 \\
505.mcf       & 2 &        12 &   1\;228 &   4\;778 &      304 &   1\;197 &      530 &   2\;149 \\
507.cactuBSSN & 102 &      345 &   5\;691 & 123\;596 &   1\;251 &  24\;770 &   2\;665 &  60\;339 \\
525.x264      & 24 &       35 &  10\;963 &  87\;991 &   1\;582 &  11\;922 &   2\;961 &  22\;438 \\
526.blender   & 981 &      996 &   8\;600 & 443\;034 &   1\;917 &  99\;558 &   2\;969 & 142\;950 \\
538.imagick   & 155 &      97 &  11\;195 & 154\;125 &   2\;425 &  29\;864 &   4\;336 &  53\;287 \\
544.nab       & 12 &       20 &   5\;741 &  22\;276 &   1\;509 &   5\;593 &   2\;712 &  11\;299 \\
557.xz        & 15 &       89 &   1\;448 &  18\;935 &      220 &   1\;470 &      390 &   3\;074 \\
\midrule
emacs-29.4    & 253 &      143 &  14\;085 & 260\;284 &   3\;377 &  48\;127 &   5\;367 &  77\;533 \\
gdb-15.2      & 172 &      251 &   5\;508 & 101\;443 &   1\;179 &  36\;708 &   1\;887 &  51\;718 \\
ghostscript-10.04 & 797 &  1\;116 &   7\;042 & 441\;161 &   1\;243 &  16\;665 &   2\;078 &  30\;929 \\
sendmail-8.18.1 & 89 &     115 &   3\;752 &  39\;205 &      799 &   7\;675 &   1\;453 &  14\;410 \\
    \bottomrule
    \end{tabular}
    }
    {\scriptsize \hspace*{1mm} $^\dagger$Total lines of code, excluding whitespace and comments, divided by 1000.}
\end{table}

\subsection{Analysis Configurations}
\label{sec:configurations}

\begin{figure}
\centering
\includegraphics[width=\columnwidth]{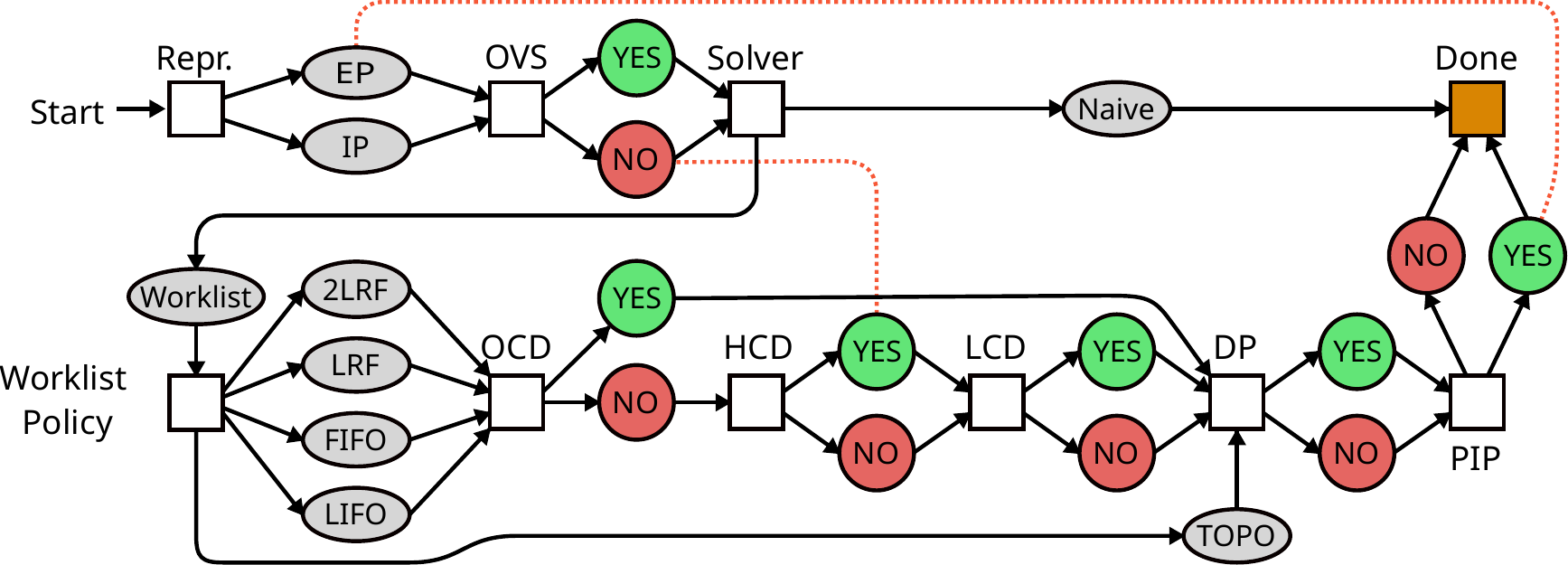}
\caption{All paths through the flowchart represent valid combinations, except for choices connected by dashed red edges, which are incompatible.
}
\label{fig:configurations-flowchart}
\end{figure}

The analysis implementation has several options that enable and disable different techniques.
Five of these techniques are from the literature, where they have been shown to improve solver performance in the context of whole-program analysis.
This paper evaluates all techniques in the context of analyzing individual files.
The full set of options is given in \autoref{tab:configurations}, and \autoref{fig:configurations-flowchart} shows a flowchart depicting all valid combinations.

\begin{table}
    \caption{The techniques that can be enabled or disabled, and the choices of solver when configuring the points-to analysis. Squares are binary choices, and dashes are exclusive choices.}%
    \label{tab:configurations}
    \resizebox{\columnwidth}{!}{
    \begin{tabular}{lll}
    \toprule  
Technique & Abbrev. & Described in \\
\midrule
\multicolumn{3}{l}{\textbf{Pointer representation}} \\
 -- Only Explicit Pointees & \explicit & \autoref{sec:external-node} \\
 -- Implicit Pointees & \implicit & \autoref{sec:implicit_pointees} \\[2mm]

\multicolumn{3}{l}{\textbf{Offline constraint processing}} \\
$\square$ Offline Variable Substitution & OVS & Rountev et al.~\cite{Rountev:PLDI2000} \\[2mm]

\multicolumn{3}{l}{\textbf{Solver}} \\
 -- Naive Solver & Naive & Andersen's thesis~\cite{Andersen:dissertation1994} \\
 -- Worklist Solver & WL & \autoref{sec:worklist} \\[2mm]

\multicolumn{3}{l}{\textbf{Worklist Iteration Order}} \\
 -- First In First Out & FIFO & \multirow{3}{*}{$\Bigg\}$ Pearce et al.~\cite{Pearce:SCAM03}}\\
 -- Last In First Out & LIFO & \\
 -- Least Recently Fired & LRF & \\
 -- 2-Phase Least Recently Fired & 2LRF & Hardekopf and Lin~\cite{Hardekopf:PLDI2007} \\
 -- Topological & TOPO & Pearce et al.~\cite{Pearce:TOPLAS2007} \\[2mm]

\multicolumn{3}{l}{\textbf{Worklist Online Techniques}} \\
$\square$ Prefer Implicit Pointees & PIP & \autoref{sec:pip} \\
$\square$ Online Cycle Detection & OCD & Pearce et al.~\cite{Hardekopf:PLDI2007} \\
$\square$ Hybrid Cycle Detection & HCD & Hardekopf and Lin~\cite{Hardekopf:PLDI2007} \\
$\square$ Lazy Cycle Detection & LCD & Hardekopf and Lin~\cite{Hardekopf:PLDI2007} \\
$\square$ Difference Propagation & DP & Pearce et al.~\cite{Pearce:SCAM03} \\
    \bottomrule
    \end{tabular}
    }
\end{table}

A configuration is described by the techniques it uses. 
For example, the configuration called \texttt{\implicit{}+WL(LRF)+OCD+PIP}
uses the implicit pointer (IP) representation, the worklist (WL) solver with the Least Recently Fired (LRF) iteration order, Online Cycle Detection (OCD), and the Prefer Implicit Pointees (PIP) techniques.
Not all combinations of choices are valid, \eg, OCD detects all cycles as soon as they appear, so there is no point in combining it with any of the opportunistic cycle detection techniques (HCD or LCD).
In total, there are \numConfigs valid configurations, and each configuration is benchmarked solving the constraint graph for each C file 50 times.
The solution is validated to ensure that all configurations produce the exact same solution.

\subsection{Implementation Details}
\label{sec:impl-details}

Constraint variables are indexed using 32-bit integers, and their $\Sol_e$ sets are implemented as hash sets.
Base constraints are placed directly into $\Sol_e$.
The six constraints added in the extended constraint language (see \autoref{tab:ip_constraints}) are
implemented as 1-bit flags on each constraint variable.
The only external library functions with special handling are \texttt{malloc}, \texttt{free}, and \texttt{memcpy}.
Cycle unification uses union-find with path compression and union-by-rank~\cite{Tarjan:Disjoint1983}, and a single $\Sol_e$ set is shared between the members of a union.
Constraints are indexed by type and node to make iterating fast.
Hash sets are used to avoid duplicated constraints.
Simple edges involving one pointer incompatible variable, \eg, $x \in M \land x \notin P$,
must be treated as pointer-integer casts.
This is achieved in practice by making $\Sol(x) \coloneq \Sol(\external)$.
When the explicit \external node is used, $x$ is unified with \external. When the implicit \external node is used, $x$ is marked both $x \sqsupseteq \external$ and $\external \sqsupseteq x$.

\section{Results}
\label{sec:results}

After performing the points-to analysis on all C files, 51\% of all pointers end up pointing to external memory (i.e., $p \sqsupseteq \external$).
Confirming that the analysis still provides useful precision is covered in \autoref{sec:precision}.
While all \numConfigs solver configurations produce the same solution, they have significant runtime variation, which is presented in \autoref{sec:runtime}.
Finally, \autoref{sec:solver-scalability} covers solver scalability.

\subsection{Precision}
\label{sec:precision}

We evaluate the precision of a points-to-analysis solution in terms of a pairwise alias-analysis client, by evaluating the load/store conflict rate, as described by Nagaraj and Govindarajan~\cite{Nagaraj:CGO2015}.
For each store instruction, the analysis is queried about possible aliasing with every other load and store instruction in the same function. The analysis returns \emph{NoAlias} if the instructions have distinct points-to sets. Otherwise, \emph{MayAlias} is returned.

For comparison, LLVM's BasicAA is used, which performs ad-hoc IR traversals to find the origin(s) of pointers.
It does not handle function calls or nested pointers, but knows that local variables that never have their address taken never alias with anything. It also tracks pointer offsets when possible.
Both analyses return \emph{MustAlias} when the pointers are identical.

\begin{figure}
    \centering
    \includegraphics[width=\columnwidth]{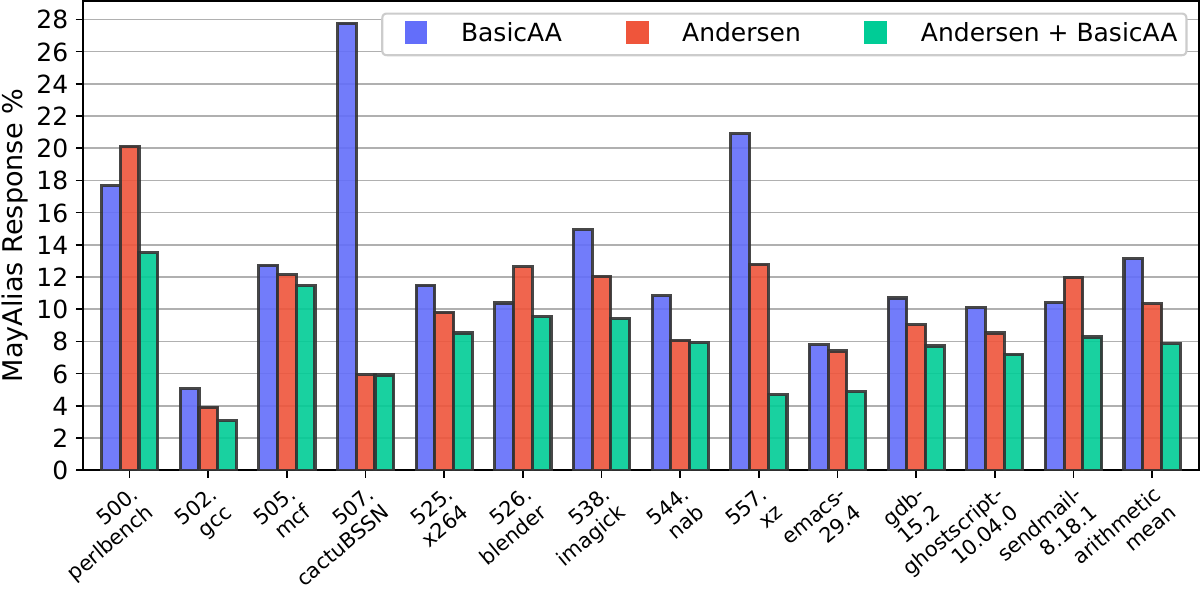}
    \caption{Percentage of intra-procedural alias analysis queries that return ``May Alias'', when querying all load/store and store/store pairs. Lower is better.}
    \label{fig:precision}
\end{figure}

\looseness -1
The results are shown in \autoref{fig:precision}. The analyses have different strengths, so the \emph{Andersen + BasicAA}-bar shows the precision achieved by combining them, as is often done in practice.
While the benefit of adding the Andersen-style points-to analysis varies by benchmark, the reduction in \emph{MayAlias}-responses is on average 40\%, indicating that the analysis adds valuable information for compiler transformations to exploit.

\subsection{Solver Runtime}
\label{sec:runtime}

The choice of solver configuration has a large impact on the runtime of the constraint-solving phase,
the most important choice being that of pointer representation.
\autoref{tab:solver-runtime} shows the distribution of solver runtimes for the benchmarked C files, for selected configurations.
It first compares the fastest configuration using explicit pointees (\texttt{EP}) against the fastest configuration using implicit pointees (\texttt{IP}) without PIP.
On average, the \texttt{EP+OVS+WL(LRF)+OCD} configuration takes 36.322\,ms per file to solve the constraint graph, while \texttt{IP+WL(FIFO)+LCD+DP} takes 2.133\,ms, around $17\times$ faster.

Next, we consider an \texttt{EP Oracle}, which always picks the fastest \texttt{EP} configuration for each file.
The oracle saves time on most files, but ends up being just 8\% faster than \texttt{EP+OVS+WL(LRF)+OCD} in total, as it is unable to significantly improve the runtime of the slowest outliers.
The \texttt{IP} configuration is thus still $15\times$ faster than the \texttt{EP Oracle}.

\begin{table}
    \centering
    \caption{Constraint graph solver runtime for selected configurations.}%
    \label{tab:solver-runtime}
    
    \resizebox{\columnwidth}{!}{
    \setlength{\tabcolsep}{3pt}
    \begin{tabular}{lrrrrrrr}
\toprule
 & \multicolumn{7}{c}{Solver Runtime [\unit{\micro\second}]} \\
 \cline{2-8}
Configuration & p10 & p25 & p50 & p90 & p99 & Max & Mean \\
\midrule
\texttt{EP+OVS+WL(LRF)+OCD}    &       40 &      168 &   1\;060 &  29\;480 & 414\;729 & 43\;437\;029 &  36\;322\\
\texttt{EP Oracle}             &       21 &      118 &      886 &  25\;191 & 347\;921 & 39\;594\;566 &  32\;376 \\
\texttt{IP+WL(FIFO)+LCD+DP}    &       19 &       62 &      249 &   2\;562 &  20\;015 & 1\;112\;770 &   2\;133  \\
\midrule
\texttt{IP+WL(FIFO)}           &       15 &       51 &      219 &   2\;337 &  19\;526 & 40\;869\;977 &  15\;370 \\
\texttt{IP+WL(FIFO)+PIP}       &       16 &       52 &      222 &   2\;260 &  14\;220 & 203\;850 &   1\;105 \\
\bottomrule
\end{tabular}
}
\end{table}

While \texttt{IP+WL(FIFO)+LCD+DP} is the fastest configuration without PIP,
the fastest configuration overall is \texttt{IP+WL(FIFO)+PIP},
shown in the bottom half of \autoref{tab:solver-runtime}.
With an average solver runtime of 1.105\,ms per file, it is 1.9$\times$ faster than the best configuration without PIP.
Adding any of the techniques from the literature to this configuration only increases the average solver runtime.
Taking advantage of these techniques would thus require heuristics to determine for which files each technique should be used.

Lastly, we consider the effect of \texttt{PIP} alone by removing it from the fastest configuration, leading to \texttt{IP+WL(FIFO)}.
As seen in \autoref{tab:solver-runtime}, enabling PIP decreases the average solver runtime by $14\times$, demonstrating that it is essential for reducing runtime. While the combination \texttt{LCD+DP} decreases runtime as well, it only reduces the
average by $7\times$ as depicted in \autoref{tab:solver-runtime}.

\begin{figure}
    \centering
    \includegraphics[width=\columnwidth]{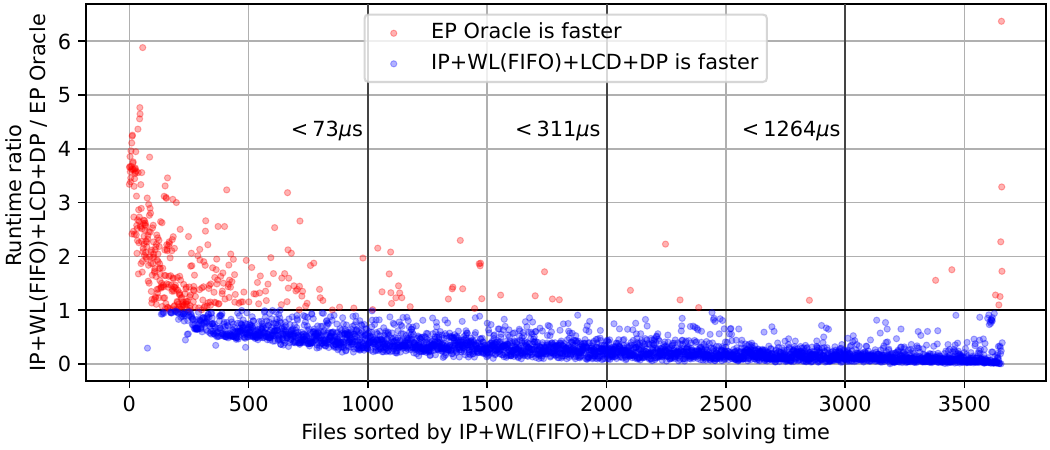}
    \medskip
    \includegraphics[width=\columnwidth]{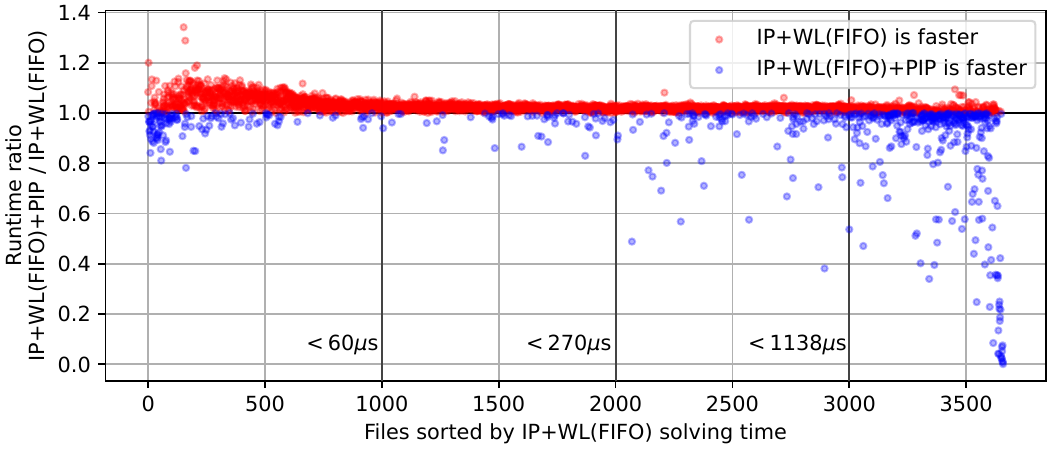}
    \caption{Solver runtime ratio of IP vs. EP (top), and PIP vs. no PIP (bottom).}%
    \label{fig:solver-runtime-ratio}
\end{figure}

\vspace{5mm}
A detailed per-file comparison is shown in \autoref{fig:solver-runtime-ratio}.
The upper graph depicts the relative solving time between the \texttt{EP Oracle} and the fastest configuration without PIP.
Among the files where the \texttt{EP Oracle} is better, 98\% are solved using the \texttt{Naive} solver, which happens to be efficient on some files regardless of pointer representation.
For the seven rightmost red dots, the \texttt{EP Oracle} uses \texttt{OVS}, as that technique proves beneficial on some large files, but not enough to be included in the on average fastest \texttt{IP} configuration.

The relative effect of enabling PIP is shown in the lower graph of \autoref{fig:solver-runtime-ratio}.
The results are not as conclusive as for \texttt{IP} vs. \texttt{EP}, but they clearly show a reduction in solving time for the longest running cases.
For 83\% of the files, enabling PIP makes the solving time slower, by 2.9\% on average.
For the files that take more than 1\,138\,\unit{\micro\second} to solve, which represent 81\% of total solver runtime, 52\% of them are slower with \texttt{PIP}, with the largest slowdown being 9\%.
In contrast, for the 48\% of the files where \texttt{PIP} is faster, the average solving time is reduced by 16$\times$.
In the most extreme case, the slowest file with \texttt{IP+WL(FIFO)}, \texttt{base/gdevp14.c} from \emph{ghostscript}, goes from 41\,s to 9.5\,ms, making an otherwise pathological file unnoteworthy.
This has a drastic effect on the Max column in \autoref{tab:solver-runtime}.
The fastest configuration without PIP, \texttt{IP+WL(FIFO)+LCD+DP}, uses 939\,ms on the same file, which is much better than 41\,\unit{\second}, but still $4.6\times$ slower than the slowest file solved with \texttt{IP+WL(FIFO)+PIP}.

\subsection{Solver Scalability}
\label{sec:solver-scalability}

The worklist algorithm spends most of its time iterating through and propagating $\Sol_e$ sets between nodes, which makes the total number of explicit pointees a relevant metric for understanding solver runtime.
They also account for the majority of memory usage.
All configurations produce identical solutions, but may use different amounts of explicit pointees.
Statistics about the number of explicit pointees produced by the different configurations are shown in \autoref{tab:explicit-pointees-statistics}.

The topmost configuration detects all cycles, which helps to reduce the number of explicit pointees by making nodes in cycles share $\Sol_e$ sets.
The next configuration uses the implicit pointee representation to represent that pointers of unknown origin may target any externally accessible memory locations.
Then comes the fastest configuration without PIP, which adds some cycle detection. Lastly, the fastest configuration overall, which uses PIP to avoid doubled-up pointees.
The results make it clear that cycle elimination is in no way a substitution for using the implicit pointee representation. It also shows that some files end up with solutions containing almost exclusively doubled-up pointees, which PIP is able to skip, reducing memory usage drastically.

\begin{table}
\centering
    \caption{Number of explicit pointees in the solutions.}%
    \label{tab:explicit-pointees-statistics}
    \resizebox{\columnwidth}{!}{
    \setlength{\tabcolsep}{3pt}
    \begin{tabular}{lrrrrrrr}
\toprule
& \multicolumn{7}{c}{Number of explicit pointees} \\
\cline{2-8}
Configuration & p10 & p25 & p50 & p90 & p99 & Max & Mean \\
\midrule
\texttt{EP+OVS+WL(LRF)+OCD}    &       38 &      305 &   3\;169 & 106\;575 & 1\;599\;946 & 154\;866\;262 & 147\;841 \\
\texttt{IP+WL(FIFO)}           &       18 &       63 &      276 &   2\;357 &  30\;162 & 2\;145\;215 &   3\;188 \\
\texttt{IP+WL(FIFO)+LCD+DP}    &       18 &       63 &      274 &   2\;323 &  29\;486 & 1\;897\;247 &   2\;816 \\
\texttt{IP+WL(FIFO)+PIP}       &       17 &       59 &      258 &   1\;977 &  11\;700 &  95\;195 &      922 \\
\bottomrule
\end{tabular}
}
\end{table}

\section{Related Work}
\label{sec:related_work}

The handling of incomplete C programs has not received much attention in the points-to analysis literature, even though the problem was already pointed out by Hind~\cite{Hind:PASTE2001} more than 20 years ago.
Most of the previous works on points-to analysis sidestep the problem by assuming whole-program analysis is performed~\cite{Lei:SAS2019, Barbar:CGO2021, Barbar:OOPSLA2021, Hardekopf:PLDI2007, Nasre:CGO2011, Pereira:CGO2009, Sui:CGO2013, Ye:SAS2014, Yu:CGO2010, Zhu:DAC2005}, where external functions, such as standard library functions or system calls, are summarized to complete the missing program parts~\cite{Pearce:SCAM03, Steensgaard:POPL1996, Lei:SAS2019, Hardekopf:PLDI2007}.

Another approach is the computation of module-based analysis summaries, that are embedded along with the compilate. The analysis consumer stitches these summaries together
to form a complete picture of the analysis results~\cite{Thakur:TOPLAS2019, Le:CC2005, Halalingaiah:OOPSLA2024, Anand:PLDI2024}.
This approach is particularly popular for JIT compiled languages, where the summaries are produced statically and then consumed by the JIT compiler, but has also been proposed for conventionally compiled languages~\cite{Rountev:CC2001, Rountev:CC2006}.
However, the problem with this approach is that it defers the analysis to link time, preventing its use by per-module optimizations.
Even at link time, a program might still not be complete, as some libraries may not have analysis results available, or libraries may be dynamically loaded at runtime.

There are only a handful of works that address the handling of incomplete C programs. Andersen~\cite{Andersen:dissertation1994} addresses the issue by introducing the abstract memory location \emph{Unknown}. The location represents all accessible memory locations at runtime, and any pointer pointing to it may alias with any other pointer in the program. This approach has the downside that a complete loss of precision arises when an unknown pointer is dereferenced on the left-hand side of an assignment.

Smaragdakis and Kastrinis~\cite{Smaragdakis:ECOOP2018} present a method for sound points-to analysis in Java.
Similarly to PIP, they avoid materializing points-to sets of pointers that can be shown to originate from what they call opaque code.
However, they model pointers originating from or escaping to opaque code as being able to point to everything, including locations that provably did not escape.
This works for Java, where all pointer accesses are based on named fields on a typed heap, but does not work in C, where pointers can be dereferenced directly.

The closest to our work is Lattner \etal~\cite{Lattner:PLDI2007}.
They present a context-sensitive and field-sensitive unification-based analysis with full heap cloning that supports incomplete programs. Their algorithm has a complete flag, denoting that all operations on objects at a node have been processed. Nodes that are reachable from unavailable external functions or global variables are not marked as complete, indicating that the information represented by this node must be treated conservatively. 
While their work yields sound solutions for incomplete programs,
it does not generalize to inclusion-based analyses, which typically give more precise solutions.

\section{Conclusion}
\label{sec:conclusion}
This paper presents an Andersen-style points-to analysis that efficiently produces sound solutions for incomplete programs.
We introduce the implicit pointee representation to efficiently represent unknown pointers that possibly target every externally accessible memory location, and show in our evaluation against several state-of-the-art techniques that it is by far the most important factor for scalability. It achieves a total speedup of 15$\times$ over always picking the fastest configuration using an explicit pointee representation.
We also introduce the Prefer Implicit Pointees (PIP) technique that further reduces the number of explicit pointees by avoiding doubled-up pointees in the solution, making the analysis an additional 1.9$\times$ faster than the fastest configuration without it. It also renders the other evaluated speedup techniques superfluous for incomplete programs, as none of them were able to outperform or aid PIP in terms of average solver runtime across all benchmarks.
Most importantly, the average solver runtime of 1.1\,ms per file makes our sound analysis practical for production compilers.

\section{Data-Availability Statement}
The implemented analysis and experimental setup, and instructions for reproducing the results in the paper, are available as an artifact on Zenodo (DOI: 10.5281/zenodo.16900791)~\cite{pip-artifact:Zenodo2025}.
The latest version of the \verb|jlm| compiler is found on GitHub~\cite{jlm:GITHUB2024}.

\ifaam
    \section*{Acknowledgments}
    PIP has partly been developed and evaluated on the IDUN/EPIC~\cite{epic:ARXIV2022} computing
    cluster at the Norwegian University of Science and Technology.
    \IEEEtriggeratref{39}
\else
    \IEEEtriggeratref{40}
\fi


\clearpage
\appendices
\section{Artifact Appendix}

\subsection{Abstract}
Our artifact provides the source code for the \verb|jlm| compiler, including an implementation of the Andersen-style analysis presented in the paper.

The artifact also includes scripts for performing the benchmarks described in the paper, and producing the figures and tables from the paper using the benchmark results.

The artifact contains the four free open-source benchmarks from \autoref{tab:benchmarks}. 
It also contains redistributable versions of the SPEC2017 benchmarks, with the exception of \verb|505.mcf|.
If you provide your own copy of SPEC2017, the full set of benchmarks from the paper will be used.

\subsection{Artifact check-list (meta-information)}

{\small
\begin{itemize}
  \item {\bf Algorithm:} Points-to analysis / Alias analysis.
  \item {\bf Program: } Four open-source benchmarks (source included). Nine benchmarks from SPEC2017 (8 of which have redistributable sources included).
  \item {\bf Compilation: } \verb|clang| 18 or above, or \verb|g++| 12 or above.
  \item {\bf Transformations: } Compiling C to LLVM IR with \verb|clang| 18.
  \item {\bf Run-time environment:}  We recommend running in a container based on the provided \verb|Dockerfile|. It can also run directly on Linux, such as Ubuntu 24.04, if the dependencies listed in the \verb|Dockerfile| are installed.
  \item {\bf Hardware: } We recommend a CPU with at least eight cores, and at least 32 GB of RAM.
  \item {\bf Execution: } Should be the sole user on CPU, with fixed frequency. Avoid running other applications while executing the artifact to limit interference.
  \item {\bf Metrics: } Analysis execution time, analysis precision.
  \item {\bf Output: } Graphs and tables included in this paper. Numbers referenced in the text of the paper. Expected results included.
  \item {\bf Experiments: } Create the Docker image and run the provided script in it.
  \item {\bf How much disk space is required (approximately)?: } 40 GB.
  \item {\bf How much time is needed to prepare workflow (approximately)?: } 5 minutes.
  \item {\bf How much time is needed to complete experiments (approximately)?: } 9 hours.
  \item {\bf Publicly available?: } Yes.
  \item {\bf Code licenses: } LGPL 2.1
  \item {\bf Data licenses: } See \verb|sources/README.md|
  \item {\bf Archived?: } \url{https://doi.org/10.5281/zenodo.16900791}
\end{itemize}

\subsection{Description}

\begin{enumerate}
\item {\it How delivered:}
Our source code, benchmarking scripts, and the four open-source benchmarks + eight redistributable SPEC benchmarks, are available at the above DOI.

\item {\it Hardware dependencies:}
The benchmarks should run on a CPU with at least eight physical cores and 32 GB of RAM.

\item {\it Software dependencies:}
We recommend using the provided \verb|Dockerfile| to build a Docker image that contains all necessary dependencies.

\item {\it Data sets:}
If you own a copy of SPEC2017, then you can provide it.
Otherwise the included redistributable sources will be used, with the following differences:
The \verb|505.mcf| benchmark is skipped. A subset of the C files in \verb|500.perlbench| are skipped. \verb|538.imagick| uses the original ImageMagick source code without SPEC's modifications. 

\end{enumerate}






\lstset{
  basicstyle=\footnotesize\ttfamily,
  columns=fullflexible,
  frame=none,
  escapeinside={(*@}{@*)}
}

\subsection{Installation}
\noindent
Extract the file \verb|pip-2026-artifact.tar.gz| in a suitable location.

\begin{lstlisting}
  $ tar xzf pip-2026-artifact.tar.gz
  $ cd pip-2026-artifact
\end{lstlisting}

If you own SPEC2017, place \verb|cpu2017.tar.xz| in the folder \verb|sources/programs/|. Avoid symlinking as it may not work inside the Docker container.

If your machine has more than 32GB of RAM and more than eight physical cores, you can update the \verb|PARALLEL_INVOCATIONS| variable in \verb|run.sh| to make the evaluation run faster. (Default is \texttt{8}).

\subsection{Experiment workflow}
\noindent
Build the Docker image:
\begin{lstlisting}
  $ docker build -t pip-2026-image .
\end{lstlisting}
Configure your CPU to run at a stable frequency, e.g., using:
\begin{lstlisting}
  $ cpupower frequency-set 
(*@\mbox{\textcolor{red}{$\hookrightarrow$}\space}@*)--min 3GHz --max 3GHz --governor performance
\end{lstlisting}
The evaluation for the paper was performed at 3 GHz, but pick a frequency that is low enough to prevent frequency boosting or throttling on your own system.

Execute the \verb|run.sh| script inside the Docker image, with the current directory mounted:
\begin{lstlisting}
  $ docker run -it
(*@\mbox{\textcolor{red}{$\hookrightarrow$}\space}@*)--mount type=bind,source="$(pwd)",target=/artifact
(*@\mbox{\textcolor{red}{$\hookrightarrow$}\space}@*)pip-2026-image ./run.sh
\end{lstlisting}
For details about what the script does, see the \verb|README.md| file.
If the command is aborted, it can be restarted, and it will continue where it left off.
To fully reset the evaluation workflow, append \verb|clean| to the end of the command.

\noindent
\textit{Running without Docker:}
If you wish to run on a different Linux system, dependencies might be located at different paths.
Compilation commands may thus need to be changed, to reference the correct include paths.
See \verb|README.md| for re-creating the list of traced compiler invocations for your own system.

\subsection{Evaluation and expected result}
\noindent
After running the experiment, results can be found in \verb|results/|:
\begin{itemize}
\item \autoref{tab:benchmarks}: {\verb|file-sizes-table.txt|}
\item \autoref{fig:precision}: {\verb|precision.pdf|}
\item \autoref{tab:solver-runtime}: {\verb|configuration-runtimes-table.txt|}
\item \autoref{fig:solver-runtime-ratio}: {\verb|ip_sans_pip_vs_ep_oracle_ratio.pdf|} and {\verb|pip_vs_best_just_without_pip_ratio.pdf|}
\item \autoref{tab:explicit-pointees-statistics}: {\verb|configuration-memory-usage-table.txt|}
\end{itemize}
The folder also contains \verb|.log| files where numbers mentioned in the text of the paper are calculated.

Results based on measured runtime will vary based on the machine,
but the overall ratios between configurations should be roughly the same.
The quantiles given in the tables should also be similarly distributed, even if they are overall faster or slower.

Precision numbers and the number of explicit pointees
can have tiny variations due to some of the open-source benchmarks configuring themselves slightly differently on different systems.

\subsection{Experiment customization}
\noindent
Custom experiments can be performed by, for example:
\begin{itemize}
\item Creating a custom \verb|sources.json| file containing compilation commands for any C program. This file can then be passed to the \verb|benchmark.py| script.
\item Modifying the \verb|benchmark.py| script to add extra flags to \verb|clang|, \verb|opt|, and/or \verb|jlm-opt|.
\item Using the \verb|jlm-opt| binary directly on any LLVM IR file made with LLVM 18, and dumping analysis runtime and/or precision metrics.
\end{itemize}
Details can be found in the \verb|README.md| file.





\begin{thebibliography}{10}
\providecommand{\url}[1]{#1}
\csname url@samestyle\endcsname
\providecommand{\newblock}{\relax}
\providecommand{\bibinfo}[2]{#2}
\providecommand{\BIBentrySTDinterwordspacing}{\spaceskip=0pt\relax}
\providecommand{\BIBentryALTinterwordstretchfactor}{4}
\providecommand{\BIBentryALTinterwordspacing}{\spaceskip=\fontdimen2\font plus
\BIBentryALTinterwordstretchfactor\fontdimen3\font minus
  \fontdimen4\font\relax}
\providecommand{\BIBforeignlanguage}[2]{{%
\expandafter\ifx\csname l@#1\endcsname\relax
\typeout{** WARNING: IEEEtran.bst: No hyphenation pattern has been}%
\typeout{** loaded for the language `#1'. Using the pattern for}%
\typeout{** the default language instead.}%
\else
\language=\csname l@#1\endcsname
\fi
#2}}
\providecommand{\BIBdecl}{\relax}
\BIBdecl

\bibitem{Surendran:CC2014}
\BIBentryALTinterwordspacing
R.~Surendran, R.~Barik, J.~Zhao, and V.~Sarkar, ``Inter-iteration scalar
  replacement using array {SSA} form,'' in \emph{Proceedings of the
  International Conference on Compiler Construction}, A.~Cohen, Ed., 2014, pp.
  40--60. [Online]. Available:
  \url{https://doi.org/10.1007/978-3-642-54807-9_3}
\BIBentrySTDinterwordspacing

\bibitem{Chitre:OOPSLA2022}
\BIBentryALTinterwordspacing
K.~Chitre, P.~Kedia, and R.~Purandare, ``The road not taken: exploring alias
  analysis based optimizations missed by the compiler,'' \emph{Proceedings of
  the {ACM} on Programming Languages}, vol.~6, p. 153:786–153:810, Oct. 2022.
  [Online]. Available: \url{https://doi.org/10.1145/3563316}
\BIBentrySTDinterwordspacing

\bibitem{Karrenberg:CGO2011}
\BIBentryALTinterwordspacing
R.~Karrenberg and S.~Hack, ``Whole-function vectorization,'' in
  \emph{Proceedings of the International Symposium on Code Generation and
  Optimization}, Apr. 2011, p. 141–150. [Online]. Available:
  \url{https://doi.org/10.5555/2190025.2190061}
\BIBentrySTDinterwordspacing

\bibitem{Lattner:PLDI2007}
\BIBentryALTinterwordspacing
C.~Lattner, A.~Lenarth, and V.~Adve, ``Making context-sensitive points-to
  analysis with heap cloning practical for the real world,'' in
  \emph{Proceedings of the {ACM} {SIGPLAN} Conference on Programming Language
  Design and Implementation}, Jun. 2007, pp. 278--289. [Online]. Available:
  \url{https://doi.org/10.1145/1250734.1250766}
\BIBentrySTDinterwordspacing

\bibitem{Landi:LOPLAS92}
\BIBentryALTinterwordspacing
W.~Landi, ``Undecidability of static analysis,'' \emph{ACM Lett. Program. Lang.
  Syst.}, vol.~1, p. 323–337, Dec. 1992. [Online]. Available:
  \url{https://doi.org/10.1145/161494.161501}
\BIBentrySTDinterwordspacing

\bibitem{Lei:SAS2019}
\BIBentryALTinterwordspacing
Y.~Lei and Y.~Sui, ``Fast and precise handling of positive weight cycles for
  field-sensitive pointer analysis,'' in \emph{SAS}, Oct. 2019, p. 27–47.
  [Online]. Available: \url{https://doi.org/10.1007/978-3-030-32304-2_3}
\BIBentrySTDinterwordspacing

\bibitem{Ye:SAS2014}
\BIBentryALTinterwordspacing
S.~Ye, Y.~Sui, and J.~Xue, ``Region-based selective flow-sensitive pointer
  analysis,'' in \emph{SAS}, M.~Müller-Olm and H.~Seidl, Eds., 2014, pp.
  319--336. [Online]. Available:
  \url{https://doi.org/10.1007/978-3-319-10936-7_20}
\BIBentrySTDinterwordspacing

\bibitem{Sui:CGO2013}
\BIBentryALTinterwordspacing
Y.~Sui, Y.~Li, and J.~Xue, ``Query-directed adaptive heap cloning for
  optimizing compilers,'' in \emph{Proceedings of the International Symposium
  on Code Generation and Optimization}, Feb. 2013, pp. 1--11. [Online].
  Available: \url{https://doi.org/10.1109/CGO.2013.6494978}
\BIBentrySTDinterwordspacing

\bibitem{Nasre:CGO2011}
\BIBentryALTinterwordspacing
R.~Nasre and R.~Govindarajan, ``Prioritizing constraint evaluation for
  efficient points-to analysis,'' in \emph{Proceedings of the International
  Symposium on Code Generation and Optimization}, Apr. 2011, pp. 267--276.
  [Online]. Available: \url{https://doi.org/10.1109/CGO.2011.5764694}
\BIBentrySTDinterwordspacing

\bibitem{Yu:CGO2010}
\BIBentryALTinterwordspacing
H.~Yu, J.~Xue, W.~Huo, X.~Feng, and Z.~Zhang, ``Level by level: making flow-
  and context-sensitive pointer analysis scalable for millions of lines of
  code,'' in \emph{Proceedings of the International Symposium on Code
  Generation and Optimization}, Apr. 2010, p. 218–229. [Online]. Available:
  \url{https://doi.org/10.1145/1772954.1772985}
\BIBentrySTDinterwordspacing

\bibitem{Pereira:CGO2009}
\BIBentryALTinterwordspacing
F.~M.~Q. Pereira and D.~Berlin, ``Wave propagation and deep propagation for
  pointer analysis,'' in \emph{Proceedings of the International Symposium on
  Code Generation and Optimization}, Mar. 2009, pp. 126--135. [Online].
  Available: \url{https://doi.org/10.1109/CGO.2009.9}
\BIBentrySTDinterwordspacing

\bibitem{Zhu:DAC2005}
\BIBentryALTinterwordspacing
J.~Zhu, ``Towards scalable flow and context sensitive pointer analysis,'' in
  \emph{Proceedings of the {ACM}/{IEEE} Design Automation Conference}, Jun.
  2005, pp. 831--836. [Online]. Available:
  \url{https://doi.org/10.1109/DAC.2005.193930}
\BIBentrySTDinterwordspacing

\bibitem{Hind:PASTE2001}
\BIBentryALTinterwordspacing
M.~Hind, ``Pointer analysis: haven't we solved this problem yet?'' in
  \emph{Proceedings of the {ACM} {SIGPLAN}-{SIGSOFT} workshop on Program
  analysis for software tools and engineering}, Jun. 2001, p. 54–61.
  [Online]. Available: \url{https://doi.org/10.1145/379605.379665}
\BIBentrySTDinterwordspacing

\bibitem{Pearce:SCAM03}
\BIBentryALTinterwordspacing
D.~Pearce, P.~Kelly, and C.~Hankin, ``Online cycle detection and difference
  propagation for pointer analysis,'' in \emph{Proceedings of the {IEEE}
  International Conference on Source Code Analysis and Manipulation}, Sep.
  2003, pp. 3--12. [Online]. Available:
  \url{https://doi.org/10.1109/SCAM.2003.1238026}
\BIBentrySTDinterwordspacing

\bibitem{Steensgaard:POPL1996}
\BIBentryALTinterwordspacing
B.~Steensgaard, ``Points-to analysis in almost linear time,'' in
  \emph{Proceedings of the {ACM} {SIGPLAN} Symposium on Principles of
  Programming Languages}, Jan. 1996, p. 32–41. [Online]. Available:
  \url{https://doi.org/10.1145/237721.237727}
\BIBentrySTDinterwordspacing

\bibitem{Hardekopf:PLDI2007}
\BIBentryALTinterwordspacing
B.~Hardekopf and C.~Lin, ``The ant and the grasshopper: Fast and accurate
  pointer analysis for millions of lines of code,'' in \emph{Proceedings of the
  {ACM} {SIGPLAN} Conference on Programming Language Design and
  Implementation}, Jun. 2007, pp. 290--299. [Online]. Available:
  \url{https://doi.org/10.1145/1250734.1250767}
\BIBentrySTDinterwordspacing

\bibitem{Barbar:CGO2021}
\BIBentryALTinterwordspacing
M.~Barbar, Y.~Sui, and S.~Chen, ``Object versioning for flow-sensitive pointer
  analysis,'' in \emph{Proceedings of the International Symposium on Code
  Generation and Optimization}, Feb. 2021, pp. 222--235. [Online]. Available:
  \url{https://doi.org/10.1109/CGO51591.2021.9370334}
\BIBentrySTDinterwordspacing

\bibitem{Barbar:OOPSLA2021}
\BIBentryALTinterwordspacing
M.~Barbar and Y.~Sui, ``Compacting points-to sets through object clustering,''
  \emph{Proceedings of the {ACM} on Programming Languages}, vol.~5, p.
  159:1–159:27, Oct. 2021. [Online]. Available:
  \url{https://doi.org/10.1145/3485547}
\BIBentrySTDinterwordspacing

\bibitem{soundiness:CACM2015}
\BIBentryALTinterwordspacing
B.~Livshits, M.~Sridharan, Y.~Smaragdakis, O.~Lhoták, J.~N. Amaral, B.-Y.~E.
  Chang, S.~Z. Guyer, U.~P. Khedker, A.~Møller, and D.~Vardoulakis, ``In
  defense of soundiness: a manifesto,'' \emph{Communications of the {ACM}},
  vol.~58, p. 44–46, Jan. 2015. [Online]. Available:
  \url{https://doi.org/10.1145/2644805}
\BIBentrySTDinterwordspacing

\bibitem{Andersen:dissertation1994}
L.~O. Andersen, ``Program analysis and specialization for the {C} programming
  language,'' Ph.D. dissertation, DIKU, University of Copenhagen, 1994.

\bibitem{ISO:Provenance6010}
\BIBentryALTinterwordspacing
{ISO/IEC JTC1/SC22}, ``Programming languages - {C} - a provenance-aware memory
  object model for {C},'' ISO/IEC, Technical Specification 6010:2025, May 2025.
  [Online]. Available: \url{https://webstore.iec.ch/en/publication/107524}
\BIBentrySTDinterwordspacing

\bibitem{SPEC-CPU2017}
\BIBentryALTinterwordspacing
{Standard Performance Evaluation Corporation}, ``{SPEC} {CPU2017} benchmark
  suite,'' 2017. [Online]. Available: \url{http://www.specbench.org/cpu2017/}
\BIBentrySTDinterwordspacing

\bibitem{Rountev:PLDI2000}
\BIBentryALTinterwordspacing
A.~Rountev and S.~Chandra, ``Off-line variable substitution for scaling
  points-to analysis,'' in \emph{Proceedings of the {ACM} {SIGPLAN} Conference
  on Programming Language Design and Implementation}, May 2000, p. 47–56.
  [Online]. Available: \url{https://doi.org/10.1145/349299.349310}
\BIBentrySTDinterwordspacing

\bibitem{Pearce:SQJ2004}
\BIBentryALTinterwordspacing
D.~J. Pearce, P.~H. Kelly, and C.~Hankin, ``Online cycle detection and
  difference propagation: Applications to pointer analysis,'' \emph{Software
  Quality Journal}, vol.~12, pp. 311--337, Dec. 2004. [Online]. Available:
  \url{https://doi.org/10.1023/B:SQJO.0000039791.93071.a2}
\BIBentrySTDinterwordspacing

\bibitem{Pearce:TOPLAS2007}
\BIBentryALTinterwordspacing
------, ``Efficient field-sensitive pointer analysis of {C},'' \emph{{ACM}
  Transactions on Programming Languages and Systems}, vol.~30, p. 4–es, Nov.
  2007. [Online]. Available: \url{https://doi.org/10.1145/1290520.1290524}
\BIBentrySTDinterwordspacing

\bibitem{Foster:Flow1997}
J.~S. Foster, M.~Fahndrich, and A.~Aiken, ``Flow-insensitive points-to analysis
  with term and set constraints,'' Tech. Rep., Jul. 1997.

\bibitem{llvm:CGO2004}
\BIBentryALTinterwordspacing
C.~Lattner and V.~Adve, ``{LLVM}: A compilation framework for lifelong program
  analysis \& transformation,'' in \emph{Proceedings of the International
  Symposium on Code Generation and Optimization}, Mar. 2004, pp. 75--86.
  [Online]. Available: \url{https://doi.org/10.1109/CGO.2004.1281665}
\BIBentrySTDinterwordspacing

\bibitem{Heintze:SIGNOTICES2001}
\BIBentryALTinterwordspacing
N.~Heintze and O.~Tardieu, ``Ultra-fast aliasing analysis using {CLA}: a
  million lines of {C} code in a second,'' \emph{{ACM} {SIGPLAN} Notices},
  vol.~36, p. 254–263, May 2001. [Online]. Available:
  \url{https://doi.org/10.1145/381694.378855}
\BIBentrySTDinterwordspacing

\bibitem{Fahndrich:PLDI1998}
\BIBentryALTinterwordspacing
M.~Fähndrich, J.~S. Foster, Z.~Su, and A.~Aiken, ``Partial online cycle
  elimination in inclusion constraint graphs,'' in \emph{Proceedings of the
  {ACM} {SIGPLAN} Conference on Programming Language Design and
  Implementation}, Jun. 1998, p. 85–96. [Online]. Available:
  \url{https://doi.org/10.1145/277650.277667}
\BIBentrySTDinterwordspacing

\bibitem{Hardekopf:SAS2007}
\BIBentryALTinterwordspacing
B.~Hardekopf and C.~Lin, ``Exploiting pointer and location equivalence to
  optimize pointer analysis,'' in \emph{Static Analysis}, H.~R. Nielson and
  G.~Filé, Eds., 2007, pp. 265--280. [Online]. Available:
  \url{https://doi.org/10.1007/978-3-540-74061-2_17}
\BIBentrySTDinterwordspacing

\bibitem{ISO:C23}
{ISO/IEC JTC1/SC22}, ``Information technology -- programming languages --
  {C},'' ISO/IEC, International Standard 9899:2024, 2024.

\bibitem{Feather:DR260}
\BIBentryALTinterwordspacing
C.~D. Feather, ``Defect report \#260,'' Sep. 2004. [Online]. Available:
  \url{https://www.open-std.org/jtc1/sc22/wg14/www/docs/dr_260.htm}
\BIBentrySTDinterwordspacing

\bibitem{Memarian:POPL2019}
\BIBentryALTinterwordspacing
K.~Memarian, V.~B.~F. Gomes, B.~Davis, S.~Kell, A.~Richardson, R.~N.~M. Watson,
  and P.~Sewell, ``Exploring {C} semantics and pointer provenance,''
  \emph{Proceedings of the {ACM} {SIGPLAN} Symposium on Principles of
  Programming Languages}, vol.~3, p. 67:1–67:32, Jan. 2019. [Online].
  Available: \url{https://doi.org/10.1145/3290380}
\BIBentrySTDinterwordspacing

\bibitem{Lee:OOPSLA2018}
\BIBentryALTinterwordspacing
J.~Lee, C.-K. Hur, R.~Jung, Z.~Liu, J.~Regehr, and N.~P. Lopes, ``Reconciling
  high-level optimizations and low-level code in {LLVM},'' \emph{Proceedings of
  the {ACM} on Programming Languages}, vol.~2, p. 125:1–125:28, Oct. 2018.
  [Online]. Available: \url{https://doi.org/10.1145/3276495}
\BIBentrySTDinterwordspacing

\bibitem{Gustedt:Provenance2021}
J.~Gustedt, P.~Sewell, K.~Memarian, V.~B.~F. Gomes, and M.~Uecker, ``A
  provenance-aware memory object model for {C},'' ISO/IEC, Draft Technical
  Specification N2676, Mar. 2021.

\bibitem{jlm:GITHUB2024}
\BIBentryALTinterwordspacing
``{JLM}: A research compiler based on the {RVSDG} {IR},'' Apr. 2024. [Online].
  Available: \url{https://github.com/phate/jlm}
\BIBentrySTDinterwordspacing

\bibitem{rvsdg:TECS2020}
\BIBentryALTinterwordspacing
N.~Reissmann, J.~C. Meyer, H.~Bahmann, and M.~Själander, ``{RVSDG}: An
  intermediate representation for optimizing compilers,'' \emph{{ACM}
  Transactions on Embedded Computing Systems}, vol.~19, pp. 49:1--49:28, Dec.
  2020. [Online]. Available: \url{https://doi.org/10.1145/3391902}
\BIBentrySTDinterwordspacing

\bibitem{Tarjan:Disjoint1983}
\BIBentryALTinterwordspacing
R.~E. Tarjan, \emph{Disjoint Sets}.\hskip 1em plus 0.5em minus 0.4em\relax
  Society for Industrial and Applied Mathematics, Jan. 1983. [Online].
  Available: \url{https://doi.org/10.1137/1.9781611970265.ch2}
\BIBentrySTDinterwordspacing

\bibitem{Nagaraj:CGO2015}
\BIBentryALTinterwordspacing
V.~Nagaraj and R.~Govindarajan, ``Approximating flow-sensitive pointer analysis
  using frequent itemset mining,'' in \emph{Proceedings of the International
  Symposium on Code Generation and Optimization}, Feb. 2015, p. 225–234.
  [Online]. Available: \url{https://doi.org/10.1109/CGO.2015.7054202}
\BIBentrySTDinterwordspacing

\bibitem{Thakur:TOPLAS2019}
\BIBentryALTinterwordspacing
M.~Thakur and V.~K. Nandivada, ``{PYE}: A framework for precise-yet-efficient
  just-in-time analyses for {Java} programs,'' \emph{{ACM} Transactions on
  Programming Languages and Systems}, vol.~41, p. 16:1–16:37, Jul. 2019.
  [Online]. Available: \url{https://doi.org/10.1145/3337794}
\BIBentrySTDinterwordspacing

\bibitem{Le:CC2005}
\BIBentryALTinterwordspacing
A.~Le, O.~Lhoták, and L.~Hendren, ``Using inter-procedural side-effect
  information in {JIT} optimizations,'' in \emph{Proceedings of the
  International Conference on Compiler Construction}, R.~Bodik, Ed., 2005, pp.
  287--304. [Online]. Available: \url{https://doi.org/10.1007/11406921_22}
\BIBentrySTDinterwordspacing

\bibitem{Halalingaiah:OOPSLA2024}
\BIBentryALTinterwordspacing
S.~Halalingaiah, V.~Sundaresan, D.~Maier, and V.~K. Nandivada, ``The {ART} of
  sharing points-to analysis: Reusing points-to analysis results safely and
  efficiently,'' \emph{Proceedings of the {ACM} on Programming Languages},
  vol.~8, p. 363:2606–363:2632, Oct. 2024. [Online]. Available:
  \url{https://doi.org/10.1145/3689803}
\BIBentrySTDinterwordspacing

\bibitem{Anand:PLDI2024}
\BIBentryALTinterwordspacing
A.~Anand, S.~Adithya, S.~Rustagi, P.~Seth, V.~Sundaresan, D.~Maier, V.~K.
  Nandivada, and M.~Thakur, ``Optimistic stack allocation and dynamic
  heapification for managed runtimes,'' \emph{Proceedings of the {ACM} on
  Programming Languages}, vol.~8, p. 159:296–159:319, Jun. 2024. [Online].
  Available: \url{https://doi.org/10.1145/3656389}
\BIBentrySTDinterwordspacing

\bibitem{Rountev:CC2001}
\BIBentryALTinterwordspacing
A.~Rountev and B.~G. Ryder, ``Points-to and side-effect analyses for programs
  built with precompiled libraries,'' in \emph{Proceedings of the International
  Conference on Compiler Construction}, Apr. 2001, pp. 20--36. [Online].
  Available: \url{https://doi.org/10.1007/3-540-45306-7_3}
\BIBentrySTDinterwordspacing

\bibitem{Rountev:CC2006}
\BIBentryALTinterwordspacing
A.~Rountev, S.~Kagan, and T.~Marlowe, ``Interprocedural dataflow analysis in
  the presence of large libraries,'' in \emph{Proceedings of the International
  Conference on Compiler Construction}, A.~Mycroft and A.~Zeller, Eds., 2006,
  pp. 2--16. [Online]. Available: \url{https://doi.org/10.1007/11688839_2}
\BIBentrySTDinterwordspacing

\bibitem{Smaragdakis:ECOOP2018}
\BIBentryALTinterwordspacing
Y.~Smaragdakis and G.~Kastrinis, ``Defensive points-to analysis: Effective
  soundness via laziness,'' in \emph{Proceedings of the European Conference on
  Object-Oriented Programming}, T.~Millstein, Ed., 2018, p. 23:1–23:28.
  [Online]. Available: \url{https://doi.org/10.4230/LIPIcs.ECOOP.2018.23}
\BIBentrySTDinterwordspacing

\bibitem{pip-artifact:Zenodo2025}
\BIBentryALTinterwordspacing
H.~R. Krogstie, ``{PIP}: Making andersen's points-to analysis sound and
  practical for incomplete c programs (artifact),'' Aug. 2025. [Online].
  Available: \url{https://doi.org/10.5281/zenodo.16900791}
\BIBentrySTDinterwordspacing

\bibitem{epic:ARXIV2022}
\BIBentryALTinterwordspacing
M.~Själander, M.~Jahre, G.~Tufte, and N.~Reissmann, ``{EPIC}: An
  energy-efficient, high-performance {GPGPU} computing research
  infrastructure,'' Feb. 2022. [Online]. Available:
  \url{http://arxiv.org/abs/1912.05848}
\BIBentrySTDinterwordspacing

\end{thebibliography}
\end{document}